\documentclass{aa}
\usepackage[varg]{txfonts}
\usepackage{graphicx}
\usepackage{tabularx}
\usepackage{amssymb}
\usepackage{enumitem}
\RequirePackage[colorlinks=true,linkcolor=blue,citecolor=blue,urlcolor=blue]{hyperref}

\begin{document}

\title{Tracing the cosmological origin of gas that fuels in situ star formation in TNG50 galaxies}
\titlerunning{The origin of gas that fuels in situ star formation}

\author{Ole Wittig\thanks{E-mail: ole.wittig@stud.uni-heidelberg.de}
\and Rahul Ramesh
\and Dylan Nelson} 
\authorrunning{O. Wittig et al.}

\institute{
Universität Heidelberg, Zentrum für Astronomie, ITA, Albert-Ueberle-Str. 2, 69120 Heidelberg, Germany \label{1}
}

\date{}

\abstract{
Based on their cosmological origin, the stars of a galaxy can be divided into two categories: those that enter through merger events (ex situ) and those born in the main progenitor (in situ). We used the TNG50 cosmological magnetohydrodynamical simulation and its Lagrangian tracer particles to explore and quantify the origin of gas that ultimately forms the in situ stars of galaxies. We tracked back the baryonic mass contributing to the $z=0$ in situ stellar populations of galaxies, studying trends in mass from dwarfs to group-scale halos. We find that more massive halos acquire this matter earlier than lower-mass halos, reflecting an overall earlier assembly of their in situ stellar mass. Defining the Lagrangian half-mass radius R$_{\rm L, 1/2}$ of a galaxy as the distance containing half of the mass that will form its in situ stars by $z=0$, we find that R$_{\rm L, 1/2}$ is larger for more massive halos at early times, reflecting larger "in situ Lagrangian regions." However, the dependence of this radius on halo mass becomes flat at $z \simeq 3$ and then inverts toward $z=0$. In addition, R$_{\rm L, 1/2}$ increases rapidly with redshift, surpassing the virial radii of halos at $z \sim 2$. This marks the cosmic epoch at which most of the gas that eventually forms the in situ stars of galaxies leaves the intergalactic medium (IGM) and enters halos, a transition that occurs earlier for more massive halos. The formation redshift of the in situ stellar component increases with halo mass, while the formation redshift of the dark matter halo decreases, indicative of a differential assembly history between these two components. Finally, we decomposed the $z=0$ in situ stellar mass into its distinct modes of accretion. Smooth accretion from the IGM is the most important for low-mass galaxies, while mergers and satellite-stripped gas become relevant and even dominant only for high-mass galaxies.
}

\keywords{galaxies: formation -- galaxies: evolution}

\maketitle


\section{Introduction}

Observations and simulations suggest that structure growth in a Lambda cold dark matter ($\Lambda\text{CDM}$) universe is hierarchical, wherein massive galaxies are formed by accreting smaller systems \citep{white78,bond96}. The detailed buildup of the stellar content of galaxies encodes information on assembly histories as well as many additional physical processes. For instance, the spatial distribution and fraction of stars that form in situ (i.e., those that form from gas in the main progenitor of their host galaxy) versus ex situ (i.e., those accreted via mergers or flyby events) reflects the star formation activity and merger events of the full progenitor tree of each galaxy across cosmic time.

Large-volume cosmological hydrodynamical simulations directly enable the study of hierarchical galaxy assembly \citep[for a recent review, see][]{vogelsberger20}. They evolve mock universes over cosmic scales and cosmic times, yielding a heterogeneous population of galaxies. To do so they account for various physical processes that impact the evolution of galaxies and their stellar content. For example, numerical models for feedback processes regulate the amount of cold gas available for star formation, typically through supernova-driven outflows in low-mass galaxies \citep{fielding17,nelson19a,pandya21} and supermassive black hole (SMBH) driven outflows in their larger counterparts \citep{oppenheimer20,davies20,ramesh23}. In addition, the galaxy gas budget is modulated by fountain flows within the halo \citep{oppenheimer08,peroux20} as well as cosmic accretion from the IGM \citep{keres05,agertz09,dekel09,nelson13}. In addition, simulated galaxies are located across a variety of large-scale environments, which in turn shapes evolutionary and internal details such as merger rates \citep{jian12,pearson24} and quenched fractions \citep{dressler80,pallero19,ayromlou21}.

As a result, the origin of the stellar content of galaxies has a large diversity across the population. In particular, the in situ fraction at $z$\,$=$\,$0$ varies from as high as $\sim$\,$95$\,\% for M$_\star$\,$\sim$\,$10^{10}$\,$\mathrm{M}_\odot$ galaxies, to as low as $\sim$\,$20$\,\% for their more massive M$_\star$\,$\sim$\,$10^{12}$\,$\mathrm{M}_\odot$ counterparts \citep[in the Illustris simulation;][]{rodriguezgomez16}. While similar qualitative trends have been noted with other simulations, results typically differ quantitatively \citep{lackner12,qu16}.

In addition, various simulation-based studies suggest that the distribution of stars follows a strong radial dependence: while the in situ component is most important in the cores and central regions of galaxies, ex situ stars are more abundant farther away \citep{lackner12, pillepich15}. This is believed to be a direct consequence of the so-called two-phase formation scenario, wherein in situ star formation dominates at early cosmic epochs when galaxies are relatively small and ex situ stellar accretion is the important factor for galaxy evolution at later times as they grow larger \citep{deluciablaizot07,oser10}.

In broad agreement with cosmological hydrodynamical simulations, semi-analytic models (SAMs) find that the stellar mass growth of ($z$\,$=$\,$0$) M$_\star$\,$\sim$\,$10^{10.8}$\,$\mathrm{M}_\odot$ galaxies is dominated by in situ star formation at $z$\,$\gtrsim$\,$1$ and mergers at later times \citep{guo08}. Similarly, $z\,=\,0$ galaxies with M$_\star$\,$\sim$\,$10^{10}$\,$\mathrm{M}_\odot$ primarily grow through star formation out to $z$\,$\sim$\,$1$ \citep{jimenez11}. On the other hand, massive ($z$\,$=$\,$0$) M$_\star$\,$\sim$\,$10^{11.5-12}$\,$\mathrm{M}_\odot$ galaxies obtain $\sim$\,$70$\,\% of their stellar content through mergers, with this ex situ growth dominating since $z$\,$\sim$\,$2$, suggesting a two-phase stellar growth scenario \citep{lee13}.

This picture is also supported by observations. In particular, \citet{perezgonzales08} used \textit{Spitzer} data to infer that more massive galaxies (M$_\star > 10^{12} \,\rm M_\odot$) tend to form their (in situ) stellar content before $z\approx 3$, whereas lower-mass systems (M$_\star < 10^{11.5}\,\rm M_\odot$) assemble half of their final mass below $z=1$. In addition, \citet{gonzalezdelgado17} observe that galactic star formation tends to peak at high redshift ($z>2$), while less massive objects have extended periods of star formation to lower redshifts. As a result, the observed in situ stellar mass fraction naturally ranges from $\sim$\,30 to 10\% for galaxies of masses $10^{11}$ to $10^{12}\,\mathrm{M}_\odot$ \citep{spavone20}, although there is significant scatter across studies \citep{spavone21}.

Observational evidence also suggests that this shift from an in situ to ex situ dominated growth period may be linked to the transformation of disk to elliptical galaxies \citep{daddi05,damjanov09,vandokkum14,vamderwal14}. Cosmological simulations have come to similar conclusions, suggesting that ex situ growth proceeds primarily through the deposition of ex situ stars at large radii by dry, minor mergers, for instance, $\mu$\,$<$\,$1/4$ or $\mu$\,$<$\,$1/10$, depending on the definition \citep{naab09,oser11,wellons16}. As a result, in situ versus ex situ stellar components provide valuable insights into the evolution and interactions of galaxies.

While many theoretical studies address the in- and ex situ stellar content of galaxies, the origin of gas that ultimately fuels in situ star formation remains unclear. For instance, these stars may form from gas contributed to by infalling satellites \citep{angalc17,wright2024}, possibly having been ejected \citep{peluso2023} or stripped \citep{mayer06,ayromlou19} prior to infall, from gas that was smoothly accreted from the IGM \citep{keres09,nelson13,mandelker20}, or from gas recycled through feedback-driven fountain flows \citep{voit15,nelson15,thompson16}.

In this paper, we use the TNG50 simulation to study the Lagrangian history of baryonic matter which forms in situ stars by $z=0$. As a benchmark for comparison, we additionally contrast results against analogous measurements for ex situ stars accreted from mergers. The paper is structured as follows: Section~\ref{sec:methods} details the IllustrisTNG simulations, as well as our analysis definitions and methodologies. Results are presented in Section~\ref{sec:results} and we summarize in Section~\ref{sec:summary}.


\section{Methods}\label{sec:methods}

\subsection{The IllustrisTNG simulations}\label{subsec:tng}

For our analysis, we derived results from the IllustrisTNG project \citep{pillepich18a,naiman18,nelson18,springel18,marinacci18}, a suite of cosmological magnetohydrodynamical simulations of galaxy formation. In particular, we focused on the highest-resolution run, TNG50-1 \citep[hereafter, TNG50;][]{nelson19a,pillepich19}. This simulates a volume of $\sim$\,($50\,\mathrm{cMpc}$)$^3$ at an average baryonic mass resolution of $\sim$\,$8 \times 10^4$\,M$_\odot$. It was run with the {\sc{Arepo}} moving-mesh code \citep{springel10}, using the IllustrisTNG model for the physics of galaxy formation and evolution \citep{weinberger17,pillepich18a}.

The TNG model includes a treatment for many key processes in galaxy formation, including star formation and evolution, stellar feedback, black hole formation, evolution and feedback, heating from a metagalactic ultra-violet (UV) background, self-shielding of dense gas, and radiative cooling, among others. These simulations solve the equations of ideal magnetohydrodynamics, starting from a primordial ($z$\,$\sim$\,$127$) seed field of $10^{-14}$\,cG \citep{pakmor11,pakmor14}, with the \cite{powell99} cleaning scheme used for divergence control.

The temporal evolution of the simulation is saved at 100 discrete instances of time, roughly logarithmically spaced between $z$\,$=$\,$20$ and $z$\,$=$\,$0$. In each of these snapshots, halos are identified using the Friends-of-Friends (FoF) algorithm with a linking length $b=0.2$ \citep{davies85}, while substructures within halos are identified using {\sc{Subfind}} \citep{springel01}. We labeled all subhalos with nonzero baryonic mass as galaxies. The galaxy that lies at the potential minimum of the halo is the central and all others are satellites. The TNG simulations adopt a cosmology consistent with the Planck 2015 analysis \citep{planckcollab16}: $\Omega_\Lambda = 0.6911$, $\Omega_{\rm m} = 0.3089$, $\Omega_{\rm b} = 0.0486$ and $h = 0.6774$.

\subsection{Classifying the cosmological origin of stars}

To classify stars based on their cosmological origin, we used the \textsc{SubLink} merger trees \citep{rodriguezgomez15}. This allowed us to connect galaxies with their progenitors and/or descendants across snapshots. Following the identification of descendants, the main progenitor was defined as the progenitor with the most massive history \citep[see also][]{deluciablaizot07}. We then labeled a star particle as in situ if it forms along the main progenitor branch (MPB) of its host galaxy \citep[following][]{rodriguezgomez16}. If not, it was labeled ex situ.

\subsection{Tracing cosmic gas flows}

To trace gas flows, we used the Monte Carlo tracer particles \citep{genel13,nelson13}. These exist within baryonic resolution elements and are exchanged probabilistically between particles and/or cells based on the mass flux between them. This scheme has been shown to improve the accuracy as compared to velocity-field tracers, particularly in regions of unordered gas motions such as shocks, turbulence, and cosmological gas accretion \citep{genel13}.  

These tracers have previously been used to understand the accretion of IGM gas into halos \citep{nelson15}, the stripping of satellite gas as they pass through their ambient backgrounds \citep{rohr23}, and to quantify the origin and evolution of cold circumgalactic gas clouds \citep{ramesh24}. This work is the first to employ them to explore the source of gas that fuels star formation.

\subsection{Gas accretion modes}\label{subsec:gas_accretion_modes}

To categorize the various origin channels, that is modes via which gas can enter a galaxy, we broadly follow \citealt{angalc17} with minor modifications. First, we defined "fresh accretion" as gas directly accreted from the IGM into the halo of interest, having never previously been bound to any other (sub)halo, and remained in the halo of interest since. 

\begin{figure*}
   \centering
   \includegraphics[width=0.50\textwidth]{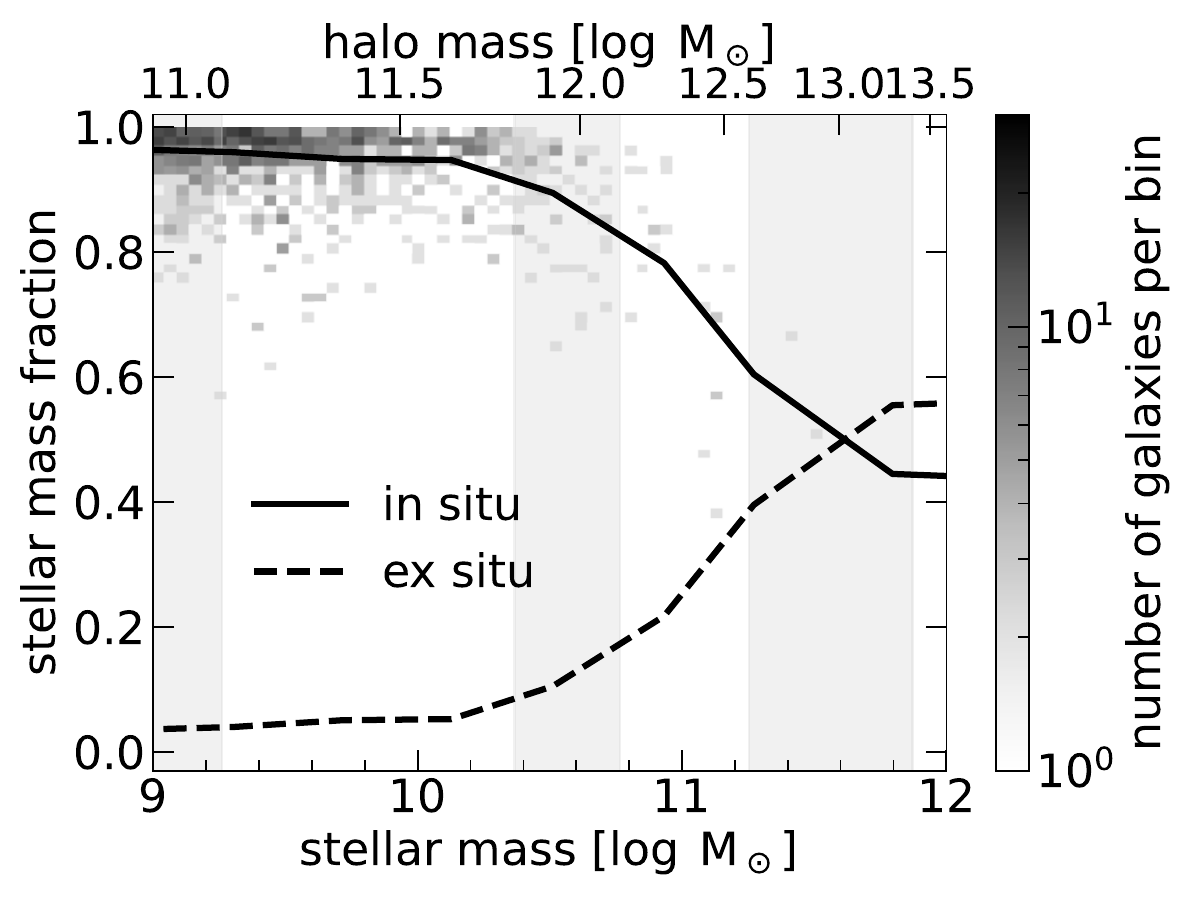}
   \includegraphics[width=0.45\textwidth]{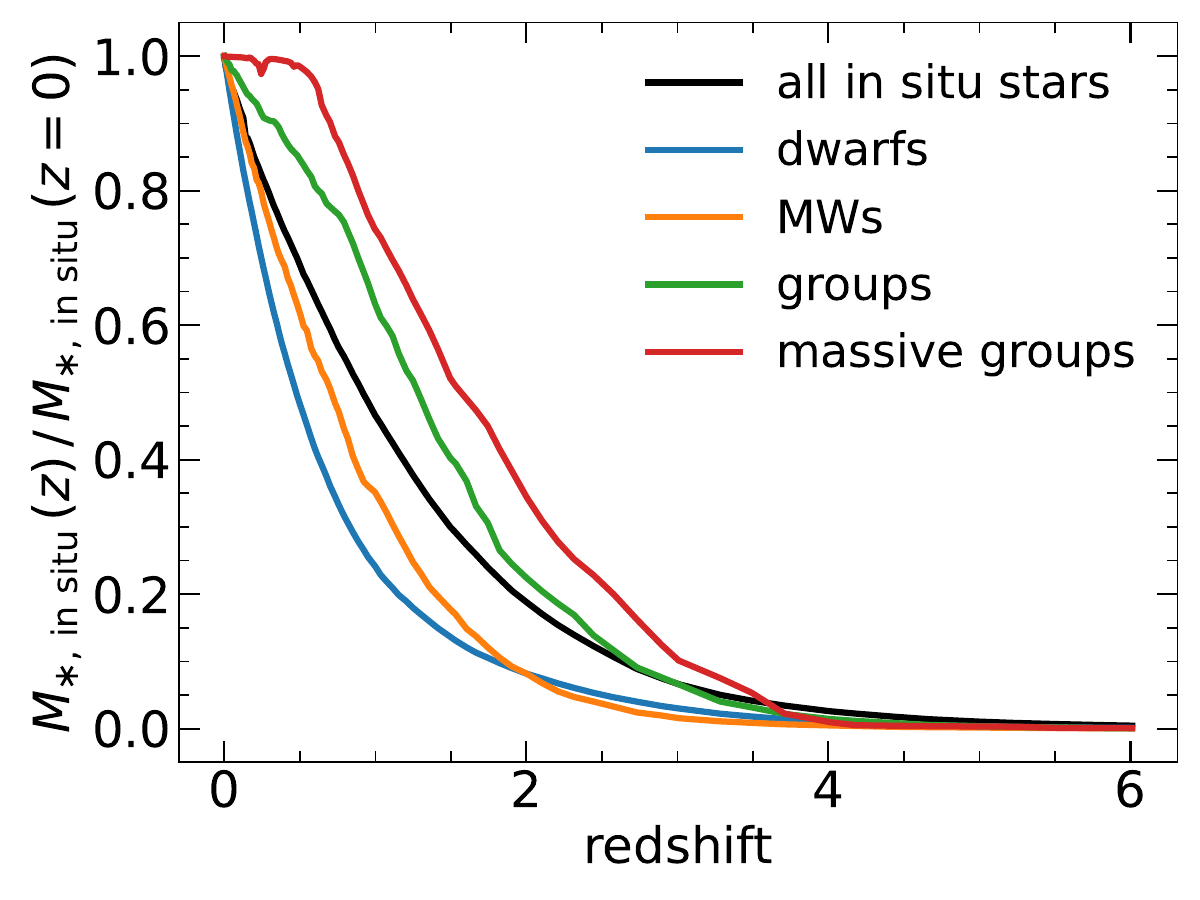}
   \caption{Distribution of in situ stellar mass fraction versus galaxy stellar mass (left) and the time evolution of this fraction (right). In the left panel, we show trends of in situ (solid) and ex situ (dashed) stellar mass fractions as a function of galaxy stellar mass at $z$\,$=$\,$0$. The in situ fraction is $\gtrsim$\,$90$\,\% at the low stellar-mass end (M$_\star$\,$\lesssim$\,$10^{10}$\,M$_\odot$), decreasing to $\sim$\,$40$\,\% for massive M$_\star$\,$\sim$\,$10^{12}$\,M$_\odot$ galaxies. For reference, the top axis shows the corresponding halo masses and the background colored pixels show the halo-to-halo scatter. The three vertical gray bands in the background show the three halo mass bins used in this work (Table~\ref{table:halo_bins}). 
   In the right panel, we show the in situ mass buildup of galaxies over cosmic time, normalized to their $z$\,$=$\,$0$ mass. The blue, orange, green, and red curves correspond to dwarfs, MWs, groups, and more massive groups, respectively. For comparison, we also include the evolutionary profile of all in situ stars from the simulation domain in black. More massive galaxies form a given fraction of in situ stellar mass at earlier times.}
   \label{fig:ex_in_situ_frac}
\end{figure*}

For the following analyses, we defined a characteristic length scale R$_{\rm SF, 1/2}$ as the comoving radius that encloses half the star-forming gas of a galaxy. Distinctly, gas can also accrete "clumpy" due to the mergers of satellites. We separated this deposition of gas by satellites (i) directly into the gaseous disks of centrals ($\leq$\,$2\,$R$_{\rm SF, 1/2}$) and (ii) gas stripped or removed at larger distances. To do so, we identified where the gas tracer is located in the snapshot directly after the merger event, that is the first snapshot when the infalling satellite no longer exists. Gas being unbound from the satellite inside of $2\, $R$_{\rm SF, 1/2}$ was labeled as "mergers", while gas removed outside of this distance is "stripped" from (infalling) satellites.

Gas can be removed from other distinct (central) halos and subsequently fall into the halo of interest. This was labeled as "stripped or ejected." Such material could be driven out of the original halo due to outflows, or dynamically or tidally stripped in a close interaction, that is a flyby or an ongoing merger. By definition, we thus only considered tracers that remain in the IGM for at least one snapshot.

Finally, gas exchanged between the galaxy and its halo, for example, due to feedback-driven outflows, was labeled as "wind recycled." In this category, we only considered gas moving from $r<0.1\,$R$_{\rm 200c}$ to $r>0.25\,$R$_{\rm 200c}$, that is from the central galaxy region into the outer halo or even beyond the halo boundary. We only include this category for better comparison with \citealt{angalc17}, as it does not contribute to the accretion of new gas.


\section{Results}\label{sec:results}

\subsection{An overview of the in situ content of galaxies} \label{subsec:insitu_distribution}

In Fig.~\ref{fig:ex_in_situ_frac}, we begin with an overview of the in situ stellar mass content of $z$\,$=$\,$0$ central galaxies. The left panel shows trends of in situ and ex situ stellar mass fractions as a function of galaxy stellar mass. For reference, we include the corresponding (mean) halo mass as the top x-axis. The solid black curve shows the median profile of stars that form in situ, while the ex situ fraction is given by the dashed black line. For the in situ case, we also visualize the statistics and scatter at fixed stellar mass with the histogram pixels and corresponding color bar.

On average, the in situ fraction is roughly constant at $\gtrsim$\,$90$\,\% for galaxies at the low-stellar mass end (M$_\star$\,$\lesssim$\,$10^{10}$\,M$_\odot$). Toward higher masses, the fraction drops monotonically to $\sim$\,$60$\,\% at M$_\star$\,$\sim$\,$10^{11}$\,M$_\odot$ and further to $\sim$\,$40$\,\% for their more massive counterparts (M$_\star$\,$\sim$\,$10^{12}$\,M$_\odot$). The ex situ fraction has inverse mass trends, increasing from as low as $\sim$\,$5$\,-\,$10$\,\% at the low-mass end (M$_\star$\,$\sim$\,$10^{10}$\,M$_\odot$), to as high as $\sim$\,$60$\,\% at M$_\star$\,$\sim$\,$10^{12}$\,M$_\odot$. The two sum to unity by definition.

These results are in good qualitative agreement with \citealt{rodriguezgomez16}, where a similar exploration of the stellar mass assembly of galaxies was performed using the Illustris simulations \citep{genel14,vogelsberger14a,vogelsberger14b,sijacki15}. They also find the ex situ fraction to be $\sim$\,$5$\,-\,$10$\,\% for galaxies less massive than M$_\star$\,$\sim$\,$10^{10}$\,M$_\odot$, but report slightly higher values of $\sim$\,$70$\,\% for M$_\star$\,$\sim$\,$10^{12}$\,M$_\odot$ galaxies. This increased ex situ star formation is likely due to weaker feedback in Illustris. This leads to galaxies that are over-massive in their stellar content in low-mass halos, ultimately contributing a greater amount of ex situ stellar material \citep[see][]{pillepich18a}. We note that the box sizes and resolutions of these two simulations also differ, possibly giving rise to additional differences due to low number statistics at the high-mass end (compare Table \ref{table:halo_bins}).

At the low-mass end, our results are also broadly consistent with the cosmological simulations used in \citealt{lackner12}, who report ex situ fractions of $\sim$\,$10$\,$-$\,$25$\,\% in $10^{10}$\,M$_\odot$\,$\lesssim$\,M$_\star$\,$\lesssim$\,$10^{11}$\,M$_\odot$ galaxies. They however find the ex situ stellar content to only account for $\sim$\,$15$\,$-$\,$35$\,\% in more massive galaxies ($10^{11}$\,M$_\odot$\,$\lesssim$\,M$_\star$\,$\lesssim$\,$10^{12}$\,M$_\odot$). This is likely due to the absence of AGN feedback in their simulations, resulting in excess in situ star formation. On the contrary, the cosmological simulations of \citealt{oser10}, which used a different prescription for feedback recipes, suggest an ex situ fraction consistent with our findings at the high-mass end, but higher by a factor of $3$\,$-$\,$5$ for low-mass galaxies. This highlights the importance of feedback physics for the stellar mass assembly of galaxies. 

\begin{figure*}
   \centering
   \includegraphics[width=0.49\textwidth]{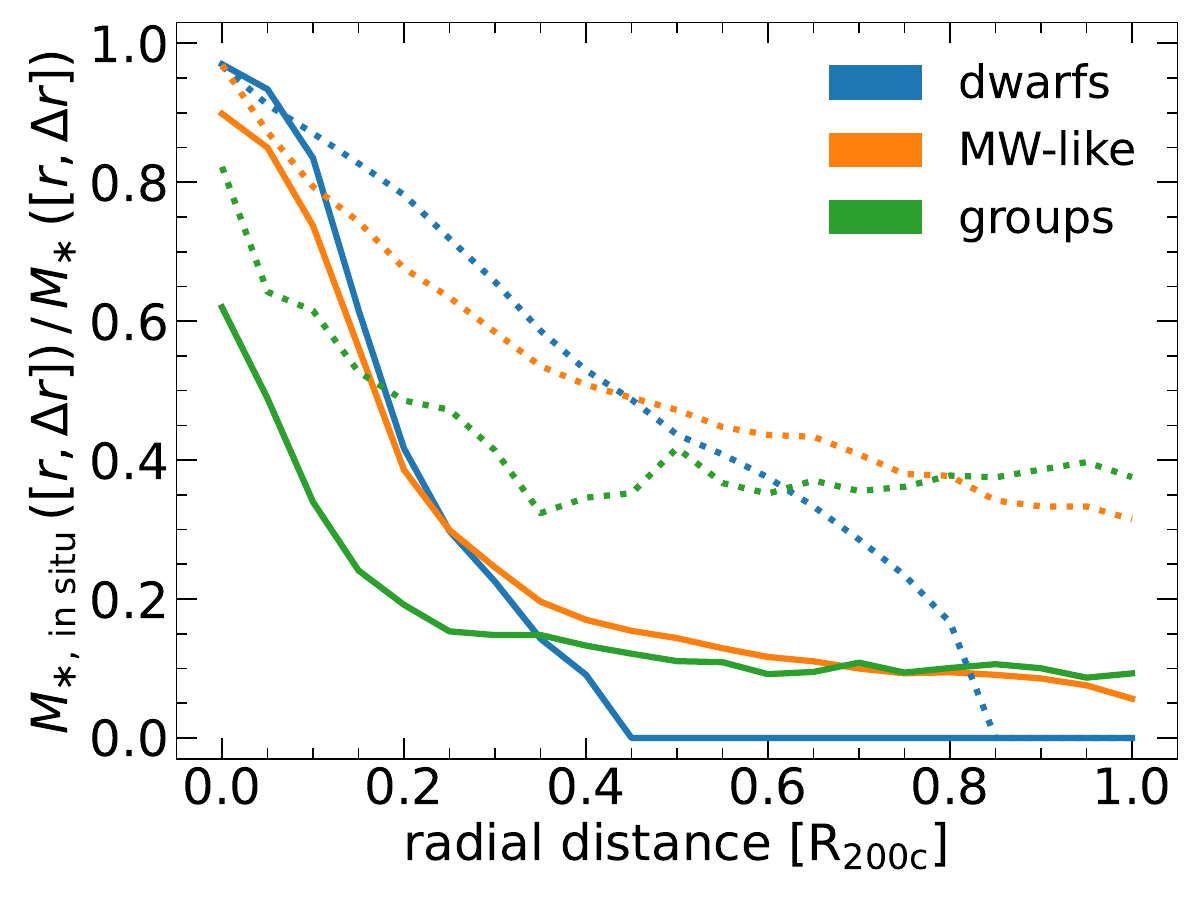}
   \includegraphics[width=0.49\textwidth]{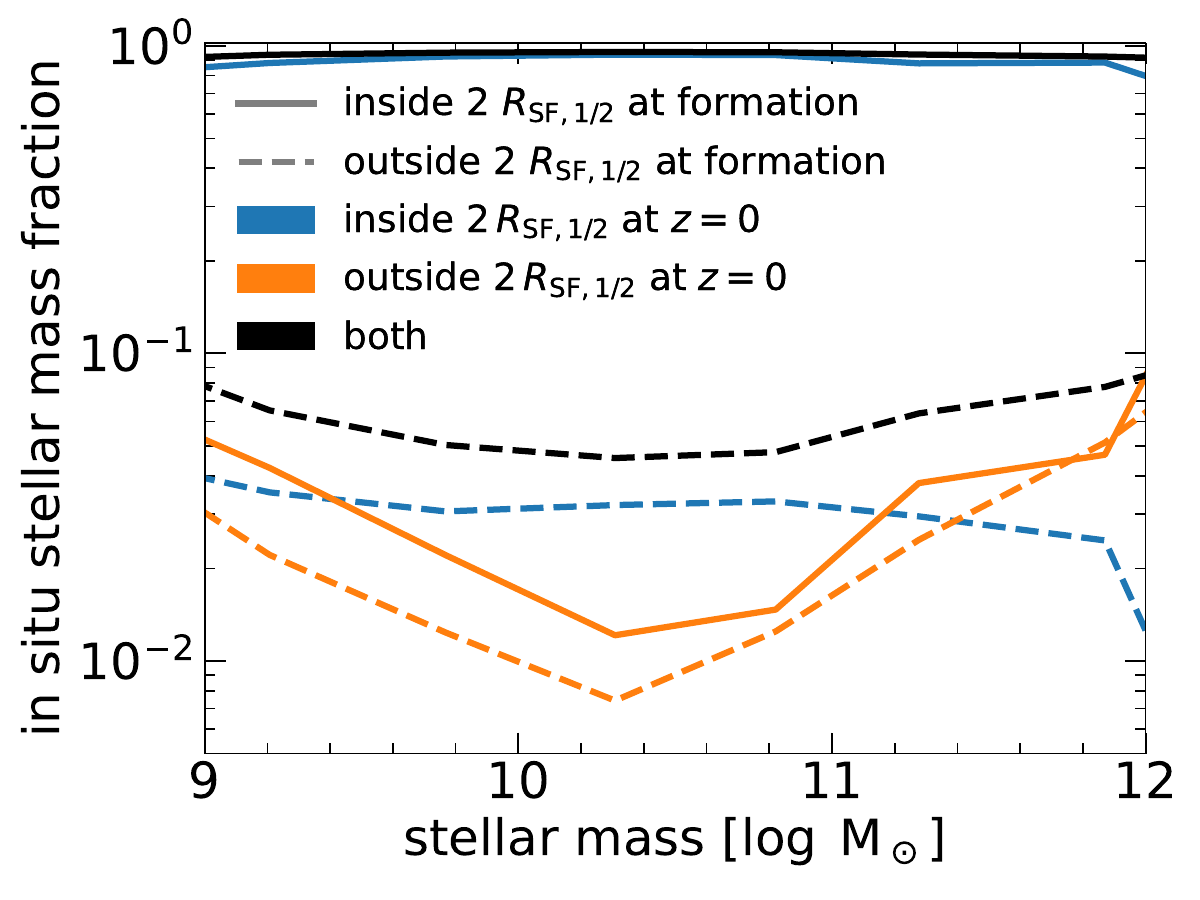}
   \caption{Radial distribution of the in situ stellar mass fraction (left) and distinguishing stellar origins (right). The left panel shows radial profiles of the in situ stellar fraction for dwarfs (blue), Milky Way mass galaxies (orange), and groups (green). Solid curves correspond to $z$\,$=$\,$0$ profiles, and dotted to the progenitors of these galaxies at $z$\,$=$\,$2$. In all three mass bins, in situ stars dominate the central regions of the halo, while ex situ stars dominate in the outskirts. This radial dependence is stronger at $z$\,$=$\,$0$. The right panel shows the in situ stellar mass fractions formed inside (solid) and outside $2$\,R$_{\rm SF, 1/2}$ (dashed), relative to the total in situ component as a function of $z$\,$=$\,$0$ galaxy stellar mass. Blue curves correspond to stars located within $2$\,R$_{\rm SF, 1/2}$ at $z$\,$=$\,$0$, those outside in orange, and the total in black. Most stars formed inside continue to reside within this region at $z$\,$=$\,$0$ following their formation, while most of the stars forming outside $2$\,R$_{\rm SF, 1/2}$ migrate inward, except in the highest mass galaxies.}
   \label{fig:insitu_profile_and_true_false}
\end{figure*}

\renewcommand{\arraystretch}{1.4}
\begin{table}[h!]
    \caption{Overview of the halo mass bins commonly used throughout our analysis.}
  \centering
  \begin{tabular}{ccc}
    \hline\hline
    Galaxy Category & Halo Mass Range & $N_{\rm halos,z=0}$\\
    \hline
    Dwarfs & $10^{10.8} \leq \mathrm{M}_\mathrm{halo} / \mathrm{M}_\odot \leq 10^{11.2}$ & 1315\\
    Milky Ways & $10^{11.8} \leq \mathrm{M}_\mathrm{halo} / \mathrm{M}_\odot \leq 10^{12.2}$  & 173  \\
    Groups & $10^{12.6}\ \leq \mathrm{M}_\mathrm{halo} / \mathrm{M}_\odot \leq 10^{13.4}$ & 40\\
    \hline
  \end{tabular}
  \tablefoot{We focus on three disjoint subsets: low-mass dwarfs, intermediate-mass Milky Way-like halos, and high-mass groups. The second column specifies their range in halo mass, while the number of halos (at $z$\,$=$\,$0$) in the corresponding bin is listed in the third column.\label{table:halo_bins}}
\end{table}
\renewcommand{\arraystretch}{1.0}

With semi-analytic models, \citealt{lee13} find that galaxies at the high-mass end ($10^{11.5}$\,M$_\odot$\,$\lesssim$\,M$_\star$\,$\lesssim$\,$10^{12.0}$\,M$_\odot$) assemble $\sim$\,$70$\,\% of their stellar content via merger accretion, roughly $\sim$\,$15$\,$-$\,$20$\,\% higher than our findings. At the low-mass end ($10^{10.5}$\,M$_\odot$\,$\lesssim$\,M$_\star$\,$\lesssim$\,$10^{11.0}$\,M$_\odot$), their predictions ($\sim$\,$20$\,\%) are however in better agreement with our findings.

In the right panel of Fig. \ref{fig:ex_in_situ_frac}, we explore the evolution of the in situ mass of galaxies over cosmic time, normalized to their $z$\,$=$\,$0$ value. The blue, orange, and green curves correspond to three galaxy subsamples used throughout this work: dwarfs, Milky Ways (MWs), and groups, respectively. As summarized in Table~\ref{table:halo_bins}, we define these to be central galaxies of halos in bins of $10^{10.8} \leq \mathrm{M}_\mathrm{200c} / \mathrm{M}_\odot \leq 10^{11.2}$, $10^{11.8} \leq \mathrm{M}_\mathrm{200c} / \mathrm{M}_\odot \leq 10^{12.2}$ and $10^{12.6}\ \leq \mathrm{M}_\mathrm{200c} / \mathrm{M}_\odot \leq 10^{13.4}$, respectively. For comparison, we also include evolutionary profiles of more massive groups ($10^{13.4}\ \leq \mathrm{M}_\mathrm{200c} / \mathrm{M}_\odot$) in red, and of all in situ stars from the simulation in black.

By definition, all profiles decline from unity at $z$\,$=$\,$0$. The in situ growth fraction of dwarfs drops to $\sim$\,$0.3$ at $z$\,$\sim$\,$1$ and further to $\lesssim$\,$0.1$ by $z$\,$\sim$\,$2$. While trends are qualitatively similar in other bins, there is a clear correlation with mass: galaxies embedded in more massive halos form a given fraction of in situ stellar mass at earlier times. For example, groups assemble $\sim$\,$50$\,\% of their ($z$\,$=$\,$0$) in situ stellar mass by as early as $z$\,$\sim$\,$1.5$. 

Interestingly, a flattening in these evolutionary profiles prior to $z$\,$=$\,$0$ is only visible in the red curve ($10^{13.4}\ \leq \mathrm{M}_\mathrm{200c} / \mathrm{M}_\odot$), that is the in situ stellar mass of these galaxies has remained roughly constant since $z$\,$\sim$\,$0.5$. \citealt{oser10} note a similar trend in their simulations, where they show the in situ star formation of massive galaxies (M$_{\rm halo} = 4.5\times10^{12} - 2.7\times10^{13}\,\mathrm{M}_\odot h^{-1}$) plateaus by $z$\,$\sim$\,$0.6$ and subsequent stellar mass growth proceeds solely through the ex situ accretion of stellar material.

This two-phase scenario of early time in situ dominated growth, followed by a late time epoch of ex situ, is absent in the other mass bins we consider. That is, the two-phase scenario holds only for the most massive galaxies. In the case of groups (green curve), a change in slope is apparent at $z$\,$\sim$\,$0.8$, signifying a gradual reduction in the level of in situ dominance. This is likely a direct consequence of the onset of strong kinetic mode black hole feedback, thereby quenching galaxies and preventing extended periods of (in situ) star formation \citep{terrazas20,zinger20,pillepich21}.

In Fig.~\ref{fig:insitu_profile_and_true_false}, we explore the radial distribution of stars throughout the halo. In the left panel, we show median spherically averaged radial profiles of the in situ stellar fraction for central galaxies in dwarf (blue), Milky Way (orange), and group (green) halos. Solid curves correspond to $z$\,$=$\,$0$ profiles, and dotted to the main progenitors of these galaxies at $z$\,$=$\,$2$.  

In all three mass bins, in situ stars dominate the central regions of the halo and fractions decay toward the outskirts. For instance, at $z$\,$=$\,$0$, roughly $\sim$\,$60$\,\% of stars at the centers of groups are in situ while this fraction is as low as $\sim$\,$10$\,\% at the virial radius (R$_{\rm{200c}}$). Furthermore, a correlation with halo mass is present: fractions are higher at the centers of lower-mass halos, consistent with Fig.~\ref{fig:ex_in_situ_frac}, and profiles are typically steeper. We note that these trends are in qualitative agreement with past studies that have inferred the centers of halos to be in situ dominated as these are the primary sites of star formation \citep{pillepich15}. Mergers typically deposit their (ex situ) mass at larger galactocentric distances \citep{amorisco17}, except for major mergers that are capable of penetrating deep into the halo \citep{rodriguezgomez16}. 

Contrasting solid versus dotted curves, we see that in situ fractions are higher at $z$\,$=$\,$2$ in all mass bins and at all distances. Stellar accretion via mergers thus grows increasingly important toward later cosmic epochs \citep{moster13,vulcani16}, even if a galaxy has not transitioned into a second phase of ex situ dominated stellar mass growth (Fig.~\ref{fig:ex_in_situ_frac}).

\begin{figure*}
   \centering
   \includegraphics[width=0.98\textwidth]{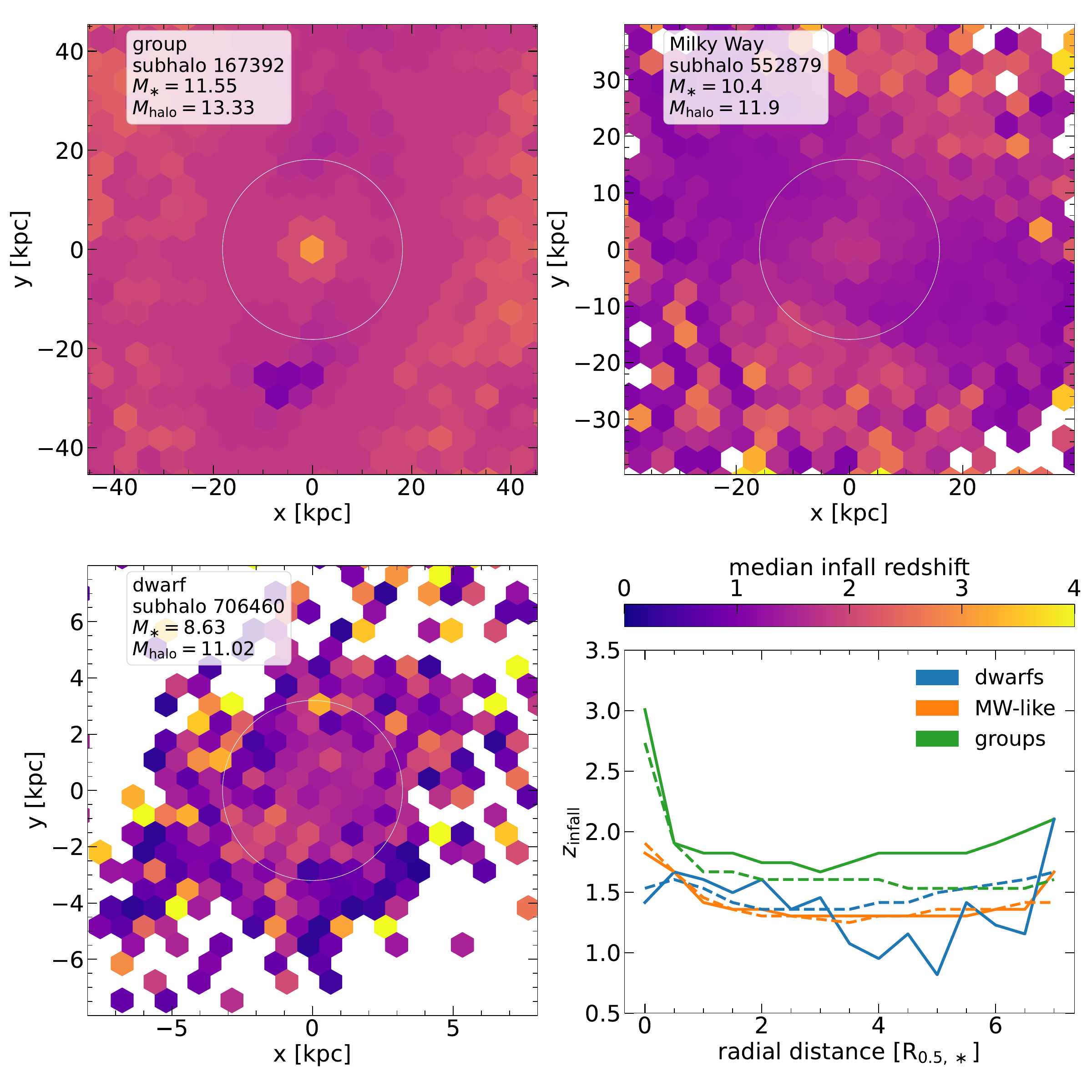}
   \caption{Visual impression of the spatial structure of the infall redshifts $z_{\mathrm{infall}}$ of tracers that reside in in situ stars at $z$\,$=$\,$0$. Representative examples of a group, Milky Way-like, and dwarf galaxy (random orientations) are shown in the upper left, upper right, and lower left panels, respectively. Colors show the median $z_{\mathrm{infall}}$ along a projection that extends through the entire halo. The lower right panel quantifies these trends with radial profiles. Solid lines show (individual) profiles of the galaxies from the three previous panels, while dashed curves the medians of all galaxies in each mass bin. Overall, the core regions of more massive galaxies are made of stars that formed from gas that entered the halo at earlier times.}
   \label{fig:mean_infall_image}
\end{figure*}

In situ stars may form throughout the gaseous halo, for example in stripped cool gas from satellites \citep{goeller23,rohr24,ahvazi24}, cool outflows \citep{maiolino17,decataldo24}, or cool clouds in the CGM \citep{canning14,nelson20}. Such stars would typically be labeled in situ, although, they are in fact a semi-distinct "mid situ" class intermediate between true in situ and ex situ. 

We explore this mechanism by examining the radial migration of in situ stars over time in the right panel of Fig.~\ref{fig:insitu_profile_and_true_false}. To identify the star-forming region of a galaxy, we defined a characteristic length scale R$_{\rm SF, 1/2}$ as the comoving radius that encloses half the star-forming gas of the galaxy. Solid curves show the fraction of stars that form within $2$\,R$_{\rm SF, 1/2}$ and those that form outside are shown through dashed lines. Curves are further split based on the galactocentric distance of stars at $z$\,$=$\,$0$: those within $2$\,R$_{\rm SF, 1/2}$ are shown in blue and those outside in orange. The sum of the blue and orange components is shown in black. 

\begin{figure}
   \centering
   \includegraphics[width=0.48\textwidth]{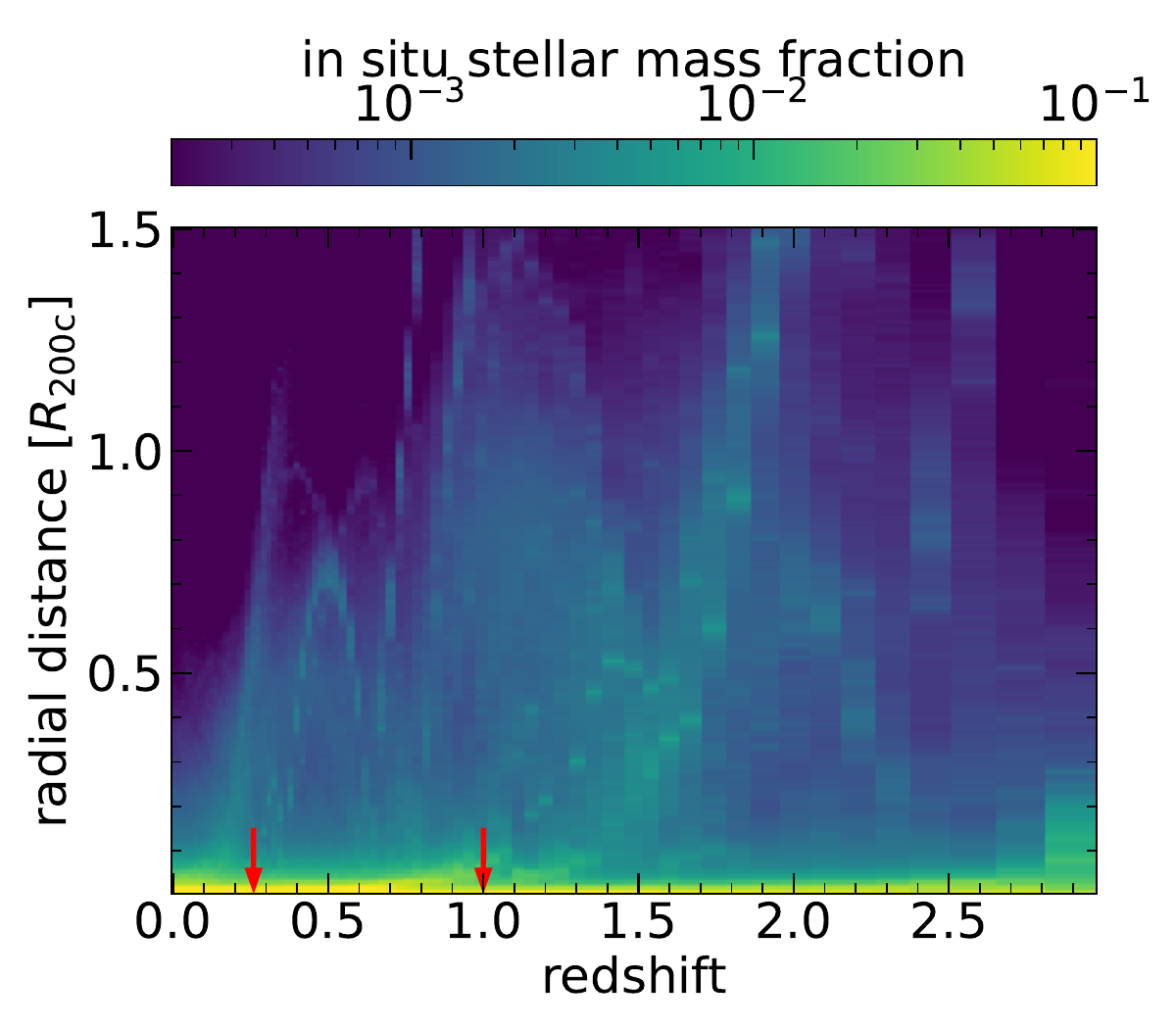}
   \caption{Temporal evolution of the radial profile of the ($z$\,$=$\,$0$) in situ mass fraction, for a single galaxy. The selected example is the same group-mass halo as in Fig.~\ref{fig:mean_infall_image}. Here we show the fraction of tracers at a given redshift in each radial bin, i.e., each column is independently normalized to unity. Red arrows mark the redshifts of major mergers within the displayed time range. While the central regions are dominated by high in situ fractions at all redshifts, more complex features are visible at larger galactocentric radii, showcasing the heterogeneous nature of gas flows and baryonic accretion.}
   \label{fig:rad_prof_sub}
\end{figure}

Across the entire range of stellar masses, most ($\gtrsim$\,$95$\,\%) in situ stars that form within $2$\,R$_{\rm SF, 1/2}$ continue to reside within the star-forming gaseous disk following their formation. These stars typically possess significant rotational support as they condense out of dense ISM gas, thereby restricting them to largely circular orbits around their sites of formation \citep{lackner12,rodriguezgomez16}. A low fraction ($\lesssim$\,$5$\,\%) of these stars may migrate to larger distances \citep{halle18,okalidis22,boecker23}, but these fractions may be overestimated if the gaseous disk size decreases over time, in particular at the high-mass end (M$_\star$\,$\gtrsim$\,$10^{10.5}$\,M$_\odot$) due to strong SMBH feedback and quenching in TNG \citep{nelson18}.

On the contrary, a subdominant fraction of stars that form outside $2$\,R$_{\rm SF, 1/2}$ remain outside the extent of the central gas disk. Most of these migrate inward toward the center following their formation \citep{elmegreen09,hartmann11}. Overall, stars forming outside $2$\,R$_{\rm SF, 1/2}$ are however rare, $\lesssim$\,$10$\,\% of the in situ population at all galaxy stellar masses.

\subsection{Case studies of individual galaxies}\label{subsec:case_study}

To motivate the analyses we focus on through the rest of the text, we begin with a case study of several individual galaxies. This also allows us to examine features that are typically washed out when multiple galaxies in a given mass bin are stacked.

In Fig.~\ref{fig:mean_infall_image}, we visualize the infall redshifts ($z_{\mathrm{infall}}$) of tracers that reside in in situ stars at $z$\,$=$\,$0$. We define this to be the redshift at which a tracer last crossed the virial radius of its $z$\,$=$\,$0$ host. Representative examples of a group (M$_\star = 10^{11.5}\,\rm M_\odot$), a Milky Way mass halo (M$_\star = 10^{10.4}\,\rm M_\odot$) and a dwarf galaxy (M$_\star = 10^{8.6}\,\rm M_\odot$) are shown in the upper left, upper right and lower left panel, respectively. All images are centered on the galaxy and colors correspond to the median $z_{\mathrm{infall}}$ along a projection that extends through the entire halo.

For the group and Milky Way-like galaxy, the central regions evidence higher $z_{\mathrm{infall}}$ (orange) than their immediate surroundings (purple). This is consistent with the inside-out galaxy formation picture \citep{pan15}. As one transitions farther away from the galaxy, the density of in situ stars drops (Fig.~\ref{fig:insitu_profile_and_true_false}) and colors are hence noisy. While some stars are formed from gas that entered the galaxy at low redshifts (blue), others clearly originate from gas that was accreted at earlier times (yellow). 

In the lower right panel, we quantify this with radial profiles of $z_{\rm infall}$. Solid lines show (individual) profiles of the galaxies from the three previous panels, while dashed curves show the medians of all galaxies in the corresponding mass bins. As seen above, the median $z_{\mathrm{infall}}$ is typically higher at the very centers of groups and MWs and decreases toward larger distances. In addition, the in situ stars at the centers of groups form from gas that is accreted into the halo at earlier cosmic epochs as compared to the other mass bins. We note however that this does not necessarily imply that these stars are older, as gas can be recycled multiple times before it eventually collapses into stars \citep{angalc17,suresh19}.

We therefore explore the accretion onto a prototypical example in more detail. Fig.~\ref{fig:rad_prof_sub} shows the temporal evolution of accreting ($z$\,$=$\,$0$) in situ mass. In particular, we compute the radial profile of the in situ mass fraction, defined for each redshift as the fraction of tracers as a function of the radial bin. Values of pixels thus add up to unity along each column, but not along each row. As before, we only consider tracers that reside in in situ stars at $z$\,$=$\,$0$. We focus on the same group central galaxy from Fig.~\ref{fig:mean_infall_image}. Red arrows mark the redshift of major mergers ($\mu$\,$>$\,$1/4$) of the galaxy \citep{rodriguezgomez17, eisert23}.

\begin{figure}
   \centering
   \includegraphics[width=0.48\textwidth]{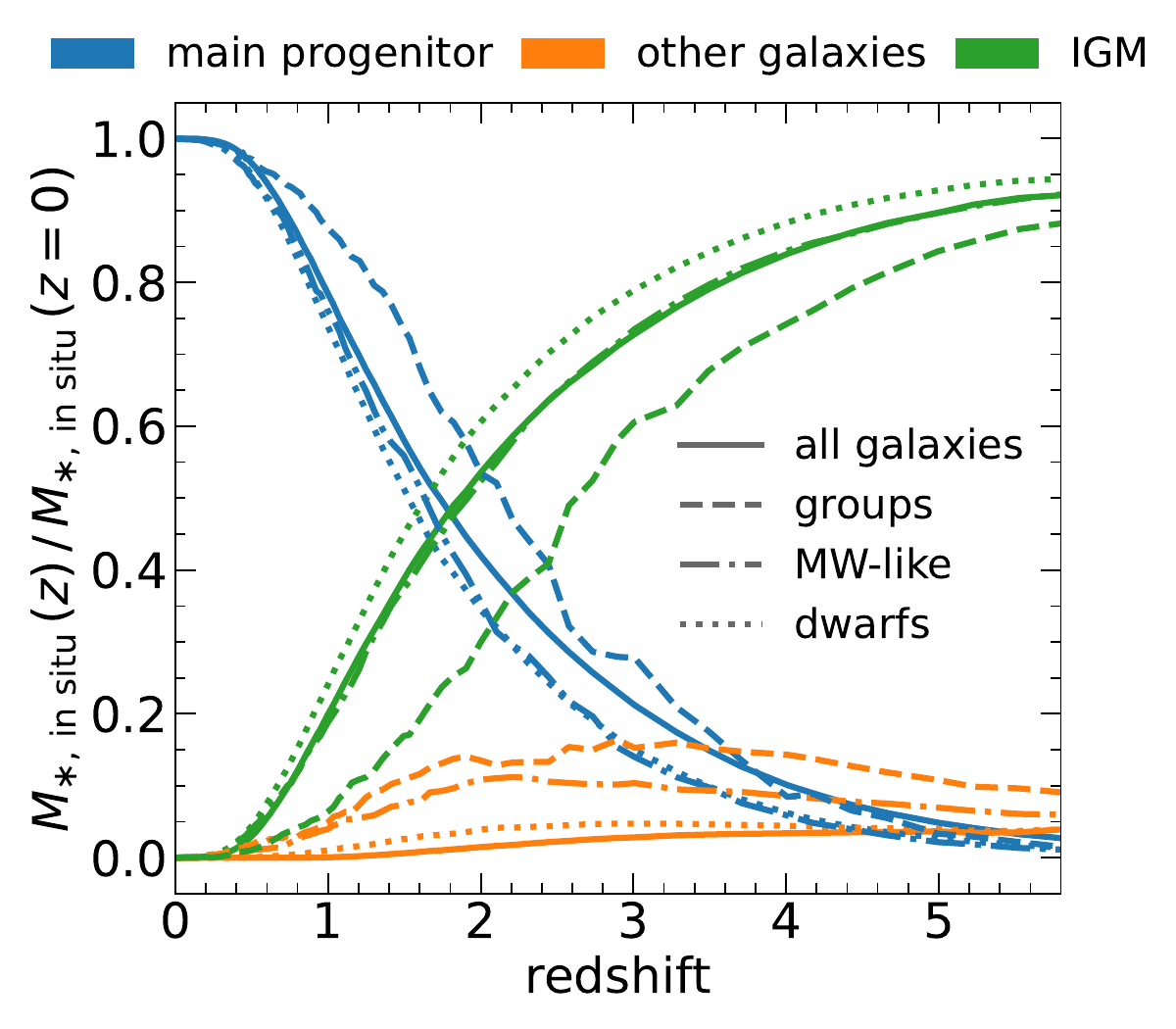}
   \caption{Fraction of the ($z$\,$=$\,$0$) in situ stellar mass of galaxies that belongs to three different origins, i.e., sources, as a function of redshift. As above, we split our sample into dwarf (dotted), Milky Way (dot-dashed), and group (dashed) halos. For comparison, in solid curves, we also include the median profile constructed from all central galaxies resolved in the simulation. The fraction of baryonic matter in the main progenitor galaxy is shown in blue, that in satellite galaxies and other halos in orange, and gas in the IGM is shown in green. A dominant fraction of gas forming in situ stars by $z$\,$=$\,$0$ was present in the IGM at high $z$, while only a low fraction was accreted in a clumpy fashion via satellites. Roughly half of the baryonic mass contributing to the $z$\,$=$\,$0$ in situ stellar mass of galaxies already resides within the main progenitor halo by $z \sim 2$.}
   \label{fig:tracer_frac_evo}
\end{figure}

Consistent with Fig.~\ref{fig:insitu_profile_and_true_false}, the central regions are dominated by high in situ fractions, at all redshifts out to $z$\,$\sim$\,$3$. Many features are simultaneously visible at larger galactocentric radii. For instance, a streak extending from the lower right to the upper left is present at $z$\,$\sim$\,$0.6$, signifying the outward motion of gas away from the galaxy and into the circumgalactic medium. In the TNG simulations, at this mass range, these are SMBH-driven outflows \citep{nelson19a} that can propagate to distances far beyond R$_{\rm{200c}}$ \citep{ayromlou23}. Most of the material expelled at $z$\,$\sim$\,$0.6$ returns to the central regions, forming an arch-like structure. In the following analysis, we classify such material as "wind recycled" (see Section \ref{subsec:gas_accretion_modes}).

Features extending along the opposite direction are also seen, that is from the upper-right toward the lower-left, as occurs for example near $z$\,$\sim$\,$0.3$. These correspond to clumpy accretion of gas as satellite galaxies fall inward. Such features are more extended at higher $z$, possibly since early-time mergers typically deposit larger amounts of cold gas \citep{cooke19,rohr24b}, thereby fueling in situ star formation at later epochs. For better understanding, the only two major mergers of the considered subhalo with mass ratios $\mu$\,$>$\,$1/4$ are marked with red arrows. They roughly agree with the visual merger redshift obtained from the streaks. While smooth accretion of IGM gas will continuously occur, it is not easily discerned here due to its low densities and mass inflow rates \citep{martizzi19,mitchell20}. 

\subsection{Statistical analysis of multiple galaxies}\label{subsec:stat_analysis}

\subsubsection{Temporal evolution of baryonic matter}\label{subsubsec:tracer_fracs}

We now undertake a statistical analysis across the galaxy population. In Fig.~\ref{fig:tracer_frac_evo}, we begin by showing the in situ stellar mass buildup as a function of redshift for central galaxies of dwarf (dashed), Milky Way (dot-dashed), and group (dashed) mass halos. For comparison, in solid curves, we also include the median profile constructed from all central galaxies resolved in the simulation. In addition, we separate the various sources of gas that fuel in situ star formation: the baryonic mass fraction in the main progenitor galaxy (blue), in satellite galaxies and other halos (orange), and in the unbound IGM (green). 

By definition, all the main progenitor curves (blue) reach unity at $z$\,$=$\,$0$. At earlier cosmic epochs, they decrease monotonically to smaller values, asymptotically approaching zero at $z$\,$\gtrsim$\,$5$. Moreover, a clear halo mass trend is visible: similar to Fig.~\ref{fig:ex_in_situ_frac}, more massive galaxies acquire a given fraction of baryons earlier than their smaller counterparts. For instance, groups attain a value of $1/2$ at $z$\,$\sim$\,$2.2$ (dashed blue), while dwarfs reach this same fraction only by $z$\,$\sim$\,$1.6$ (dotted blue). 

Qualitatively, the IGM fractions (green) have the inverse trend, both versus time and halo mass. Starting from zero at $z$\,$=$\,$0$, they increase to $\gtrsim$\,$90$\,\% by $z$\,$\gtrsim$\,$5$ and further to $100$\,\% by $z$\,$\sim$\,$12$ (not shown). Most gas fueling in situ star formation by $z$\,$=$\,$0$ is thus present in the IGM at high $z$ and is "smoothly" accreted into the halo \citep{keres05,nelson13}. Groups typically acquire a lower fraction of their gas from the IGM as compared to MWs and dwarfs.

Group-like galaxies instead accrete a slightly higher fraction of gas via gas-rich mergers (orange), which is expected given that they experience a greater overall number of mergers \citep{oser10}. On the contrary, lower-mass dwarfs generally have quieter merger histories \citep{rodriguezgomez15} and thus obtain a smaller component of their gas through this channel. However, even in the case of groups, this channel contributes a subdominant fraction ($\lesssim$\,$15$\,\%) as compared to the IGM. We also note that a nonnegligible $5-10$\% of baryons that build the in situ populations of more massive $z=0$ galaxies are already condensed into "other galaxies" even at $z \sim 6$, that is by the end of the epoch of reionization.

\begin{figure*}
   \centering
   \includegraphics[width=0.98\textwidth]{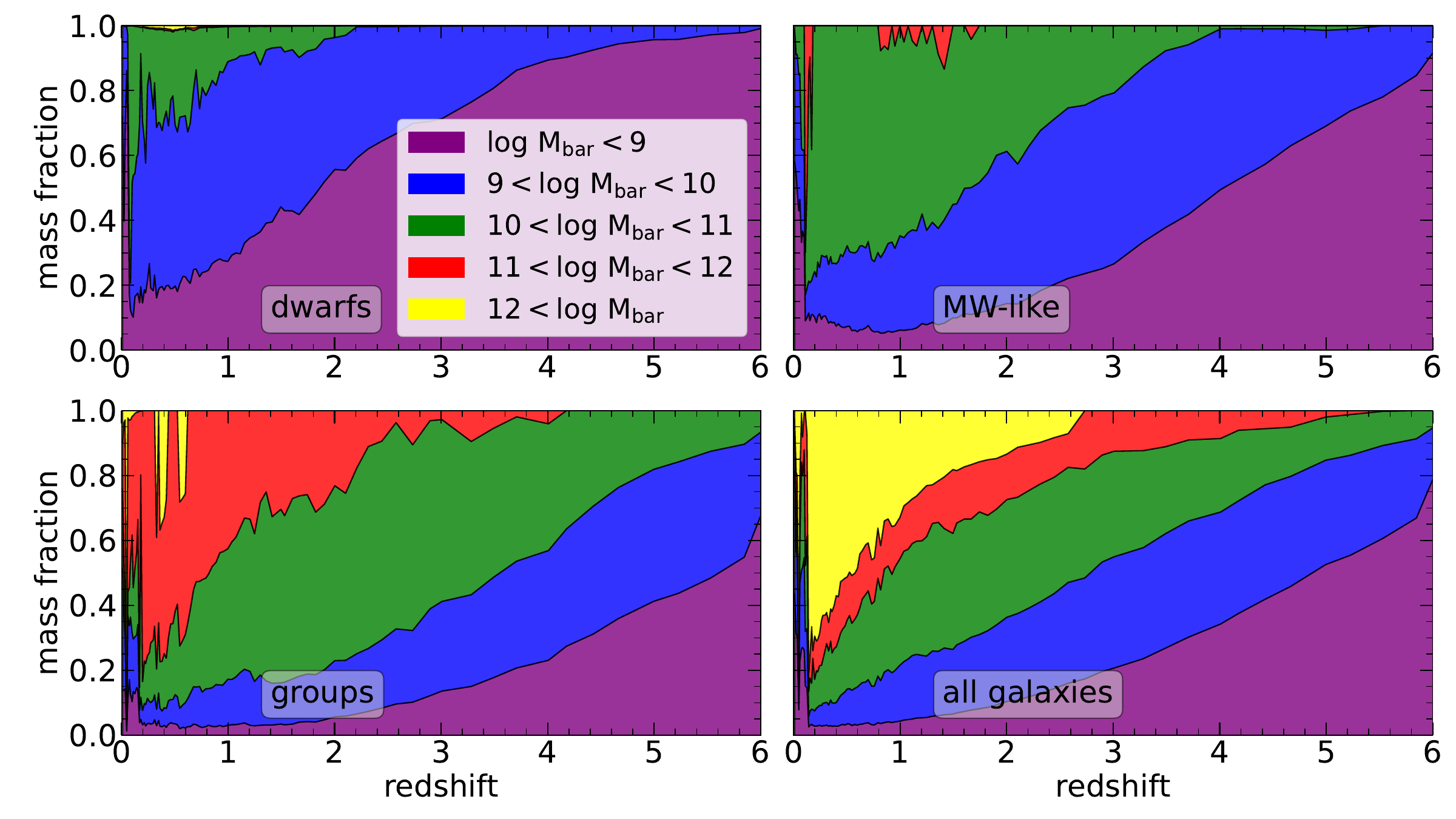}
        \caption{Fractions of the in situ stellar mass of galaxies accreted via mergers versus redshift, decomposed by the baryonic mass of the other galaxies at each redshift. In effect, we split the orange curves of Fig.~\ref{fig:tracer_frac_evo} into further components. We focus on dwarf, Milky Way, and group halos in the upper-left, upper-right, and lower-left panels, respectively. For reference, the lower-right panel shows the same analysis when combining all central galaxies resolved in the simulation. Overall, more massive centrals obtain a lower fraction of mass from under-massive structures in terms of their baryon content.}
   \label{fig:tracer_frac_evo_other_gals}
\end{figure*}

There is also significant variation from halo-to-halo (see Fig \ref{fig:tracer_frac_var_comp}). For instance, at $z$\,$\sim$\,$2$, we find the 1$\sigma$ scatter for dwarfs to be $\sim$\,$10$\,\% for baryons accreted through the IGM. A smaller scatter is present for gas obtained via mergers ($\sim$\,$5$\,\%), although the number is as large as $\sim$\,$10$\,$-$\,$15$\,\% for groups at $z$\,$\gtrsim$\,$2.5$, as expected since these massive galaxies experience a greater number of interactions.

In Fig.~\ref{fig:tracer_frac_evo_other_gals} we further decompose the baryonic contribution to the in situ stellar mass of galaxies brought in via mergers and intergalactic transfer (compare orange curves in Fig. \ref{fig:tracer_frac_evo}). In particular, we show stacked redshift trends of the fractional baryonic mass, split based on the mass of the other galaxies a tracer belongs to at each redshift. We separate these other galaxies into five bins of baryonic mass M$_{\rm bar}$, given by the sum of the total stellar plus gas mass. We focus on central galaxies of dwarf, Milky Way, and group-mass halos in the upper-left, upper-right, and lower-left panel, respectively. For reference, the lower-right panel shows analogous profiles for a stacked sample of all central galaxies resolved in the simulation. In effect, we show how galaxies of different masses contribute to the orange curves of Fig.~\ref{fig:tracer_frac_evo}.

At early cosmic epochs ($z$\,$\gtrsim$\,$3$), low-mass galaxies (M$_{\rm{bar}}$\,$\lesssim$\,$10^{10}$\,M$\odot$) dominate the fractions in all mass bins. Toward lower redshift, the contribution by more massive halos increases. In particular, more massive centrals obtain a lower fraction of mass from under-massive structures in terms of their baryon content. For instance, Milky Way-like galaxies obtain $\sim$\,$70$\,\% of their baryons via $10^{10}$\,$\lesssim$\,M$_{\rm{bar}}$\,$\lesssim$\,$10^{11}$\,M$\odot$ galaxies since $z$\,$\sim$\,$1$, but this fraction is $\lesssim$\,$30$\,\% for groups.

A low fraction of mass may also be obtained through flyby events. An instance of this is visible via the nonzero contribution by M$_{\rm{bar}}$\,$\gtrsim$\,$10^{12}$\,M$\odot$ galaxies in the top-left panel\footnote{The abrupt spikes present in all panels at $z$\,$\lesssim$\,$0.2$ are a result of relatively small tracer counts in the "other galaxies" category at low $z$ (Fig.~\ref{fig:tracer_frac_evo}). This causes sudden jumps in mass fractions when a given galaxy undergoes a major merger, that is these features are due to finite numerical resolution.}. This signature is similar to the case of "intergalactic mass transfer" \citep{angalc17}, where gas ejected by feedback from a particular halo is later accreted by other halos in its vicinity. While these M$_{\rm{bar}}$\,$\gtrsim$\,$10^{12}$\,M$\odot$ galaxies do not typically influence the gas assembly of most centrals, they do contribute significantly to cluster-mass halos. This is highlighted in the lower-right panel, which is dominated by galaxies in halos more massive than $10^{13.6}$\,M$\odot$ since we assign an equal weight to all tracers. Here, the yellow band reaches $\gtrsim$\,$60$\,$-$\,$70$\,\% at low redshift.

\subsubsection{Spatial distribution of baryonic matter}
\label{subsubsec:cumulative_radial_profiles}

\begin{figure*}
   \centering
   \includegraphics[width = 0.98\textwidth]{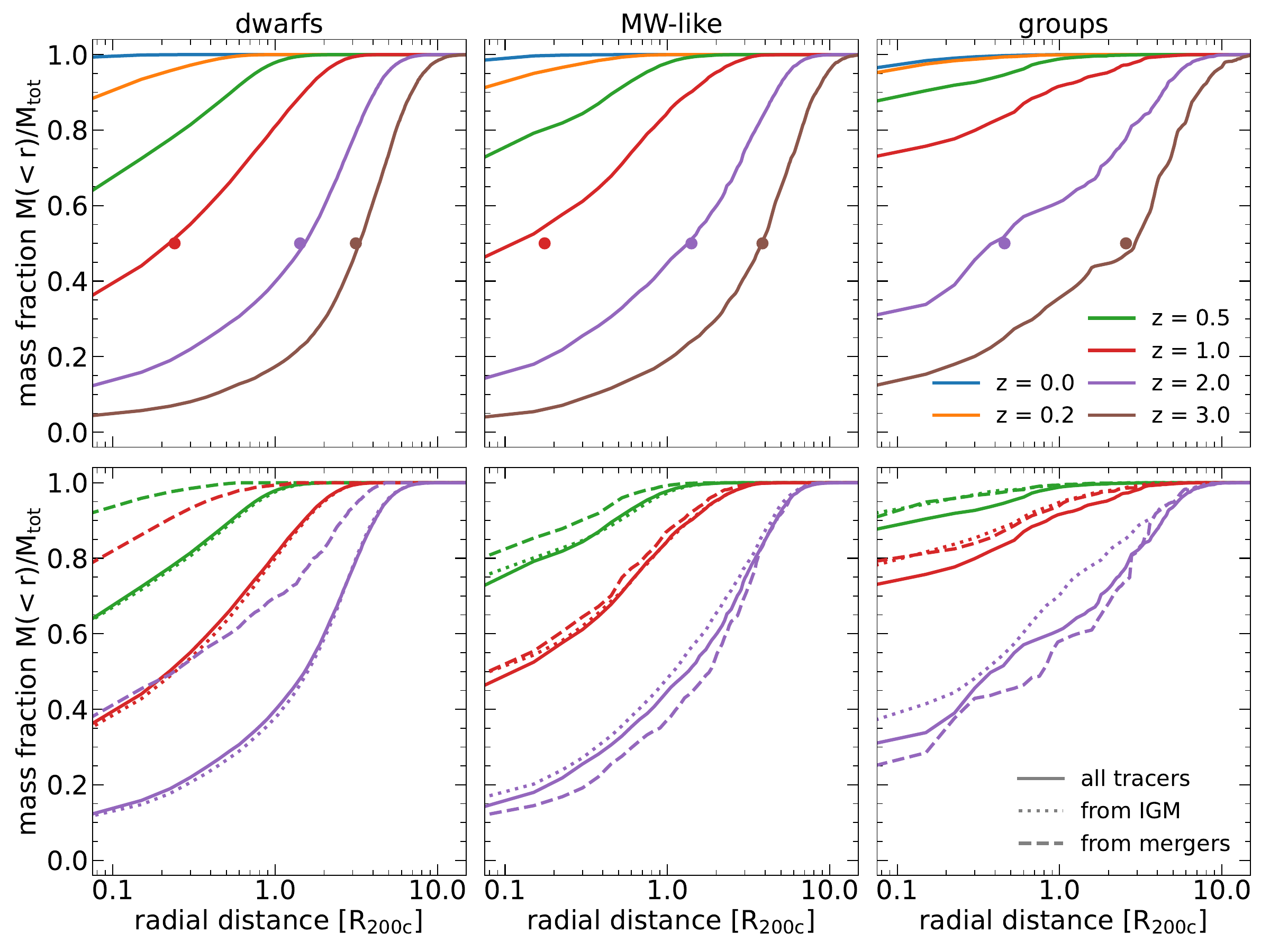}
   \caption{Radial distribution of the baryonic mass that forms the $z$\,$=$\,$0$ in situ stellar mass of dwarf (left column), Milky Way (center) and group (right) galaxies. For each radial bin, we show the total in situ baryonic mass enclosed by a sphere of this radius, normalized to the total in situ stellar mass at $z$\,$=$\,$0$. In the upper row, we display the time evolution, i.e., the collapse of this mass into galaxies with median profiles at six different redshifts. Colored dots correspond to the median galactocentric distance at which the fraction crosses a threshold of $0.5$, which we define as the Lagrangian half-mass distance ($\rm{R}_{\rm L, 1/2}$). Groups assemble their baryons earlier than lower-mass galaxies and have smaller Lagrangian half-mass distances. In the lower row, we further split a subset of curves based on their origin: baryons accreted from the IGM (dotted) versus mergers (dashed), as well as the total (solid, repeated from the top row). IGM gas collapses into halos from relatively smaller distances as compared to gas contributed by merging galaxies.}
   \label{fig:radial_profiles}
\end{figure*}

The accretion of baryonic matter into halos also leaves spatial signatures. In Fig.~\ref{fig:radial_profiles} we show radial profiles of the cumulative baryonic mass fractions of central galaxies of dwarf (left column), Milky Way (center), and group (right) mass halos. We define this fraction to be the ratio of the (tracer representative) mass within a sphere of radius $r$ ($M(<r)$) to the total in situ stellar mass at $z$\,$=$\,$0$ (M$_{\rm tot}$). In the upper row, we show median profiles at six different redshifts (colored curves). For each, the colored dot corresponds to the median galactocentric distance at which the fraction crosses a threshold of $0.5$. We label this the Lagrangian half-mass distance ($\rm{R}_{\rm L, 1/2}$). Although we use a characteristic radius to quantify this extent, the distribution of gas, particularly at high redshift, is not generally spherically symmetric \citep{natarajan00,parker07}.

At $z$\,$=$\,$0$, profiles in all three mass bins flatten to a value of unity within the virial radius, by definition. At earlier cosmic epochs, baryons constituting the $z$\,$=$\,$0$ in situ stellar population are typically spread out to larger distances. For instance, for dwarf galaxies at $z$\,$=$\,$2$, roughly $50$\,\% of these baryons reside outside of $1.5$\,R$_{\rm{200c}}$. A clear mass trend is also present: baryons are more centrally concentrated around groups, where $50$\,\% of this mass is within $0.5$\,R$_{\rm{200c}}$ at the same redshift. 

We find that the halo-to-halo variation is typically small ($\lesssim 5\%$) at $z$\,$=$\,$0$ for all mass bins (not shown), as expected since in situ stars in most galaxies populate the central regions of the halo. Toward higher redshifts, the scatter increases, with groups showing the highest level of diversity by $z$\,$=$\,$2$, implying that galaxies evolve and accrete their baryons in varied ways over time \citep{hafen19}.

\begin{figure}
   \centering
   \includegraphics[width = 0.53\textwidth]{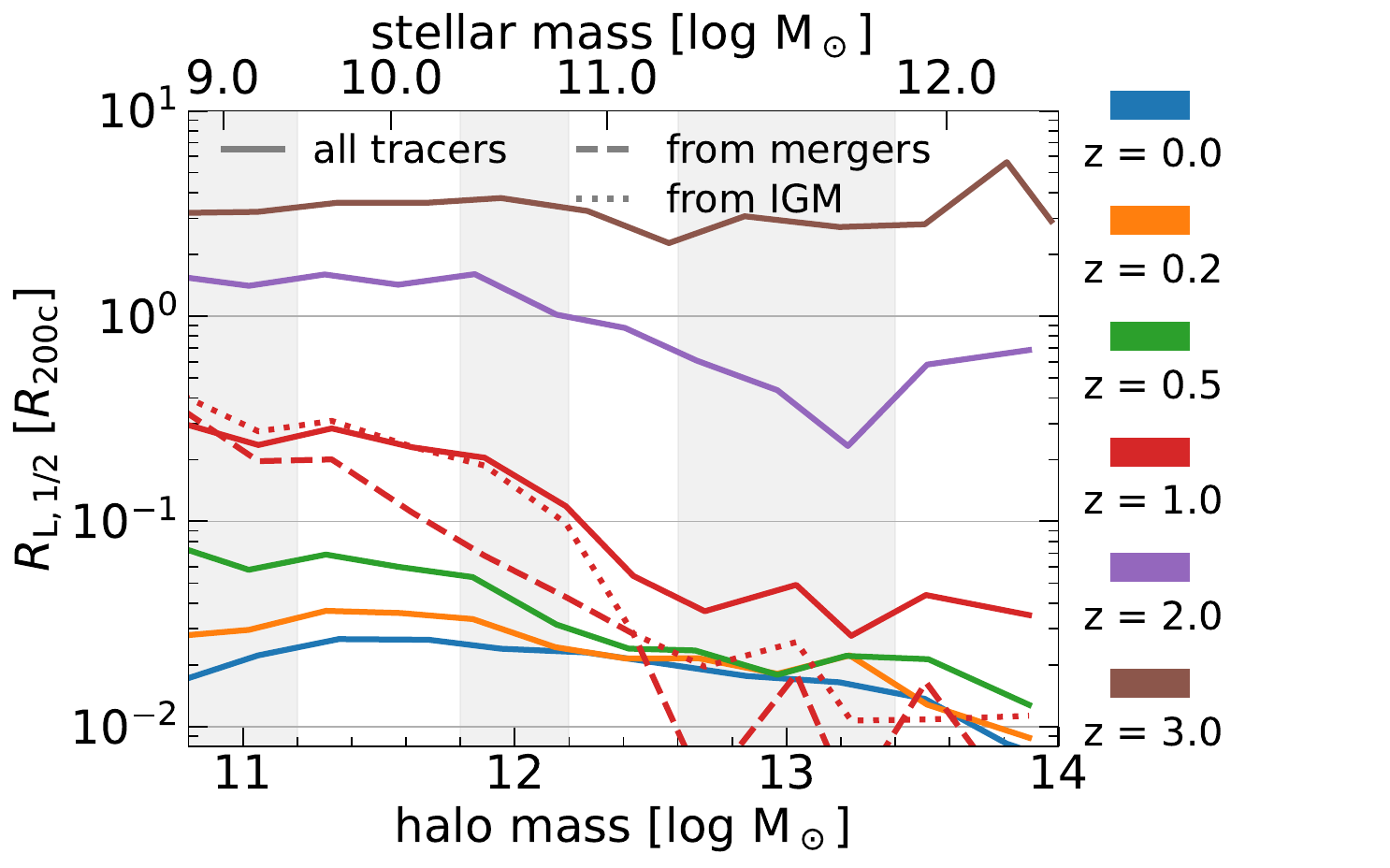}
   \caption{Trend of Lagrangian half-mass radius ($\rm{R}_{\rm L, 1/2}$) with halo mass. Solid curves are derived from all tracers that constitute the $z$\,$=$\,$0$ in situ stellar population, with colors showing six different redshifts. For $z$\,$=$\,$1$ alone, we further split based on tracers accreted from the IGM (dotted) and via mergers (dashed). For reference, the top x-axis shows the mean stellar mass as a function of halo mass. Broadly, $\rm{R}_{\rm L, 1/2}$ decreases for more massive halos, reflecting smaller relative Lagrangian regions.}
   \label{fig:hmr_vs_halo_mass}
\end{figure}

In the lower row, we split a subset of curves based on origin: baryons accreted from the IGM are shown with dotted curves and those via mergers with dashed curves. For reference, profiles from the upper panel, that is those constructed from all tracers irrespective of origin, are shown through solid curves.

At $z$\,$=$\,$2$, for Milky Way and group mass halos, fractions are higher for the IGM origin as compared to that of mergers. Gas accreted from the IGM is thus less spread out at high redshift as compared to the contribution by other galaxies. The biggest disparity between these two categories occurs for dwarfs, where gas accreted via mergers is significantly closer to the halo as compared to gas originating from the IGM.

Toward lower redshift, the offset between these two categories grows smaller for MWs and groups, reaching parity for groups by $z$\,$\lesssim$\,$1$. On the contrary, for dwarfs, IGM gas is more spread out to farther distances as compared to gas in other halos, even at these relatively low redshifts. This trend may be primarily driven by the relatively higher fraction of IGM gas contributing to in situ stars by $z$\,$=$\,$0$ in these low-mass halos (Fig.~\ref{fig:tracer_frac_evo}).

Similar to the upper row, the halo-to-halo variation is small at $z$\,$=$\,$0$, but increases toward higher redshifts. For instance, at $z$\,$=$\,$2$, the 1$\sigma$ scatter is $\sim$\,$30$\,\% for MWs and groups. This value is slightly higher for dwarfs ($\sim$\,$40$\,\%) which is likely due to overall lower rates of mergers and thus increased variability between halos. Overall, these evolutionary in situ mass profiles show that more massive galaxies assemble their baryons earlier than their low-mass counterparts, in agreement with Fig.~\ref{fig:tracer_frac_evo}.

To quantify this further, Fig.~\ref{fig:hmr_vs_halo_mass} explores the correlation of $\rm{R}_{\rm L, 1/2}$ with halo mass. Solid curves show medians derived from all tracers that constitute the $z$\,$=$\,$0$ in situ stellar mass of the central galaxy, with colors showing lines from six different redshifts. For $z$\,$=$\,$1$ alone (red), we further split based on tracers accreted from the IGM (dotted) and via mergers (dashed).

At $z$\,$=$\,$0$, the median profile anticorrelates weakly with halo mass: starting from $\sim$\,$0.025$\,R$_{\rm{200c}}$ at M$_{\rm{200c}}$\,$\sim$\,$10^{11}$\,M$_\odot$, it decreases to $\lesssim$\,$0.025$\,R$_{\rm{200c}}$ for more massive halos (M$_{\rm{200c}}$\,$\sim$\,$10^{13}$\,M$_\odot$). At all halo masses, in situ stars thus typically constitute the innermost regions of the halo, in agreement with the left panel of Fig.~\ref{fig:insitu_profile_and_true_false}. For the most massive halos (M$_{\rm{200c}}$\,$\sim$\,$10^{14}$\,M$_\odot$), $\rm{R}_{\rm L, 1/2}$ drops even further to $\lesssim$\,$0.01$\,R$_{\rm{200c}}$, as in situ star formation in these galaxies is almost entirely quenched at late times (right panel of Fig.~\ref{fig:ex_in_situ_frac}).

Toward higher redshift, this anticorrelation grows stronger. For instance, at $z$\,$=$\,$0.5$, $\rm{R}_{\rm L, 1/2}$ increases to $\sim$\,$0.07$\,R$_{\rm{200c}}$ for dwarf-mass systems (M$_{\rm{200c}}$\,$\sim$\,$10^{11}$\,M$_\odot$) while it is relatively unchanged for more massive halos (M$_{\rm{200c}}$\,$\gtrsim$\,$10^{12.5}$\,M$_\odot$) with respect to their $z$\,$=$\,$0$ value. This is a direct result of low-mass systems forming significant amounts of in situ stars at low redshift, as opposed to higher-mass halos that primarily grow through the ex situ accretion of stars \citep{oser10}. Confounding this simple picture, however, gas-rich mergers may drive the production of in situ stars in massive halos, albeit temporarily \citep{cooke19}.

At $z$\,$=$\,$1$, the anticorrelation peaks and curves begin to flatten out once again toward $z$\,$\gtrsim$\,$3$. At these early cosmic epochs, baryons constituting the $z$\,$=$\,$0$ in situ stellar population are thus typically $\gtrsim$\,$3$\,R$_{\rm{200c}}$ away from the halo, largely independent of halo mass. In addition, a 1$\sigma$ variation $\sim$\,$0.5$\,dex is present at high redshift and regardless of halo mass (not shown) implying a diversity of baryonic accretion histories.

Contrasting the IGM versus merger origins at $z$\,$=$\,$1$ shows that $\rm{R}_{\rm L, 1/2}$ is larger for gas accreted smoothly via the IGM, in agreement with Fig.~\ref{fig:radial_profiles}. Once again, a nonnegligible scatter is present at all halo masses ($\sim$\,$1$\,dex; not shown) implying large diversity across halos at fixed mass.

\subsubsection{Formation and infall times}
\label{subsubsec:formation_infall_times}

\begin{figure}
   \centering
   \includegraphics[width=0.45\textwidth]{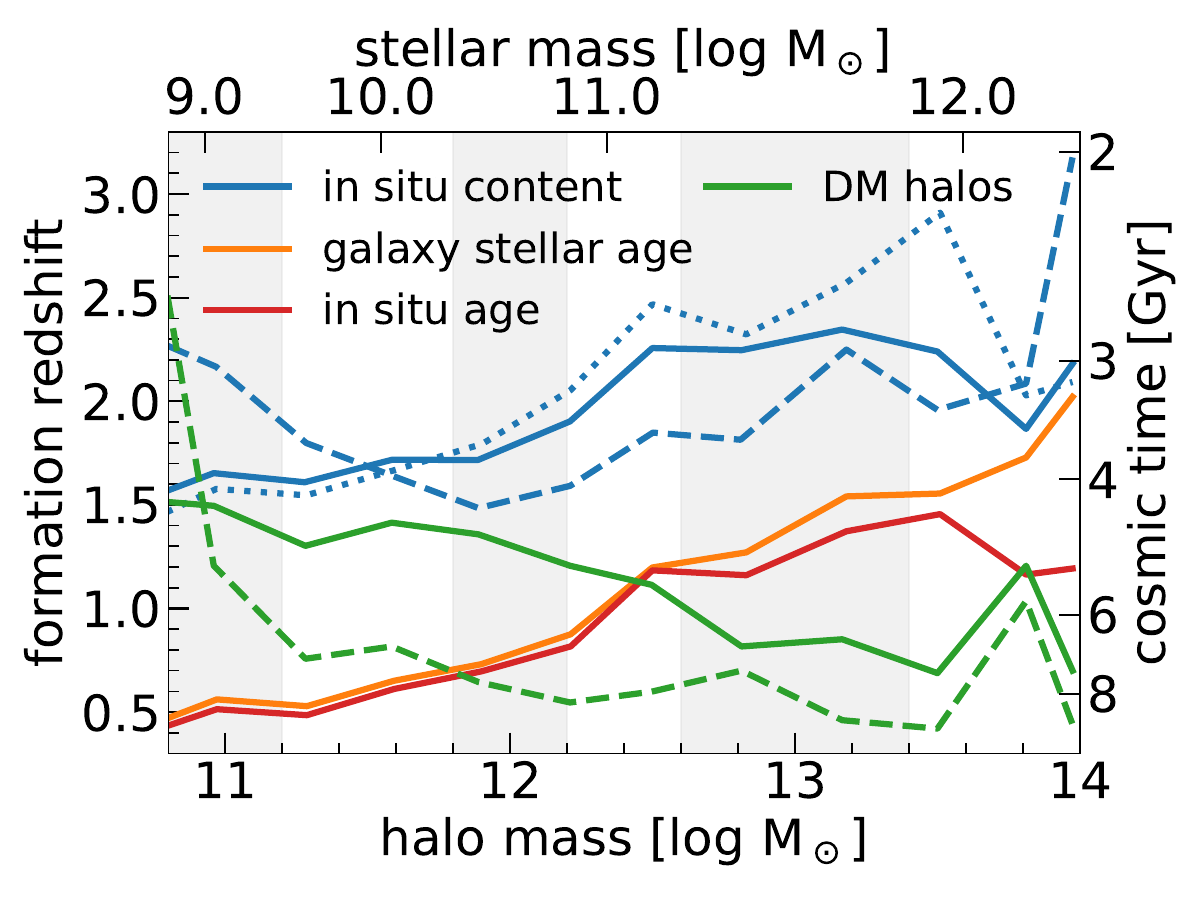}
   \caption{Formation redshift of the baryons that form the in situ stellar populations $z$\,$=$\,$0$ galaxies (blue). We also compare to the formation redshifts of the dark matter halo (solid green) and the dark matter halo core (dashed green). Solid curves show all origins, while the dashed (dotted) blue line shows mass accreted via mergers (the IGM). For comparison, we also show the mean age of all (in situ) stars of the central galaxy in orange (red). While the global dark matter (DM) halo as well as its core assemble earlier for less massive objects, the inverse holds for the stellar population, as well as the accretion of baryons that later fuel in situ star formation.}
   \label{fig:halo_insitu_formation_times}
\end{figure}

Fig.~\ref{fig:halo_insitu_formation_times} measures the assembly time of the baryonic matter fueling in situ star formation by $z$\,$=$\,$0$ (blue curve), computed as the redshift at which $\rm{R}_{\rm L, 1/2}$ equals the virial radius of the halo. We contrast the mass accreted through all origins (solid blue) to IGM origin (dotted) and merger origin (dashed) baryons.

We contrast these measurements against the formation times of the dark matter (DM) halo (solid green) and the DM core (dashed green). The former is defined as the redshift at which the halo mass M$_{\rm{200c}}$ attains half its $z$\,$=$\,$0$ value. For the latter, we instead consider the DM mass within twice the stellar half-mass radius, as it evolves. Finally, we also show the mean age of all stars of the central galaxy (orange) and the mean age of in situ stars only (red).

In agreement with past studies \citep[e.g.,][]{LiMoGao08}, we find that the DM halo typically assembles earlier for less massive objects: the formation redshift is $z$\,$\sim$\,$1.5$ for halos of mass M$_{\rm{200c}}$\,$\sim$\,$10^{11}$\,M$_\odot$, while more massive halos (M$_{\rm{200c}}$\,$\sim$\,$10^{13}$\,M$_\odot$) assemble half their DM mass only by $z$\,$\sim$\,$0.8$. The formation redshift of the DM core has a similar qualitative behavior. Except for low-mass halos (M$_{\rm{200c}}$\,$\lesssim$\,$10^{11}$\,M$_\odot$), the core typically assembles at somewhat later cosmic epochs.

The formation time of gas-building in situ stars is however qualitatively different. In particular, high-mass galaxies accrete these baryons earlier than their lower-mass counterparts. For instance, considering all origins together (solid blue line), this "formation redshift" is $z\sim$\,$1.5$ for dwarf mass halos (M$_{\rm{200c}}$\,$\sim$\,$10^{11}$\,M$_\odot$), but $z$\,$\sim$\,$2.0$ for groups and more massive objects (M$_{\rm{200c}}$\,$\gtrsim$\,$10^{13}$\,M$_\odot$). This highlights a qualitative difference between the assembly of a dark matter halo versus its baryonic counterpart.

The interpretation of this differential trend is not trivial. In the simplest scenario, we would expect the baryonic content to follow the assembly of its dark matter halo host. If the baryon accumulation is delayed with respect to the DM, for instance, due to the additional hydrodynamical forces acting against gravity, this could produce a lower formation redshift for baryons versus DM, in disagreement with our findings in Fig. ~\ref{fig:halo_insitu_formation_times}. However, DM is deposited at large galactocentric radii, leading to a minimal physical mass gain of halos \citep{wetzel15}. In addition, rapid accretion of cold gas at high redshift could lead to rapid baryonic assembly \citep[e.g.,][]{keres05, nelson15}. The link between the accretion rate of dark matter, the accretion rate of gas, and the star-formation rate and thus in situ stellar mass production of a galaxy has complex dependencies on both redshift and mass \citep{genel13}.

For halos more massive than M$_{\rm{200c}}$\,$\sim$\,$10^{11.5}$\,M$_\odot$, we find that the formation redshifts are larger (smaller) for gas accreted via the IGM (mergers), that is they accrete a bulk of their IGM gas prior to epochs where gas-rich mergers begin to dominate. Interestingly, the opposite holds for low-mass halos (M$_{\rm{200c}}$\,$\lesssim$\,$10^{11.5}$\,M$_\odot$): the late time in situ star formation of these objects is fueled by the smooth accretion of gas from the IGM. On the one hand, this crossover between the IGM (dashed blue) and merger (dotted blue) curves at M$_{\rm{200c}}$\,$\sim$\,$10^{11.5}$\,M$_\odot$ may be a consequence of the environment in which these galaxies are embedded. Dwarfs are typically isolated at late times \citep{kraljic18, laigle18, hoosain24}, implying that gas accretion from the IGM dominates in late epochs. In contrast, mergers only contribute to the early evolution. In addition, toward lower redshifts, high-mass halos develop a strong virial shock that slows down the smooth accretion of IGM, while clumpy accretion via satellites is largely unimpacted \citep{dekel09,nelson15}. This could potentially lead to higher formation redshifts for IGM-origin versus merger-origin gas for massive objects.

\begin{figure*}
   \centering
   \includegraphics[width=\textwidth]{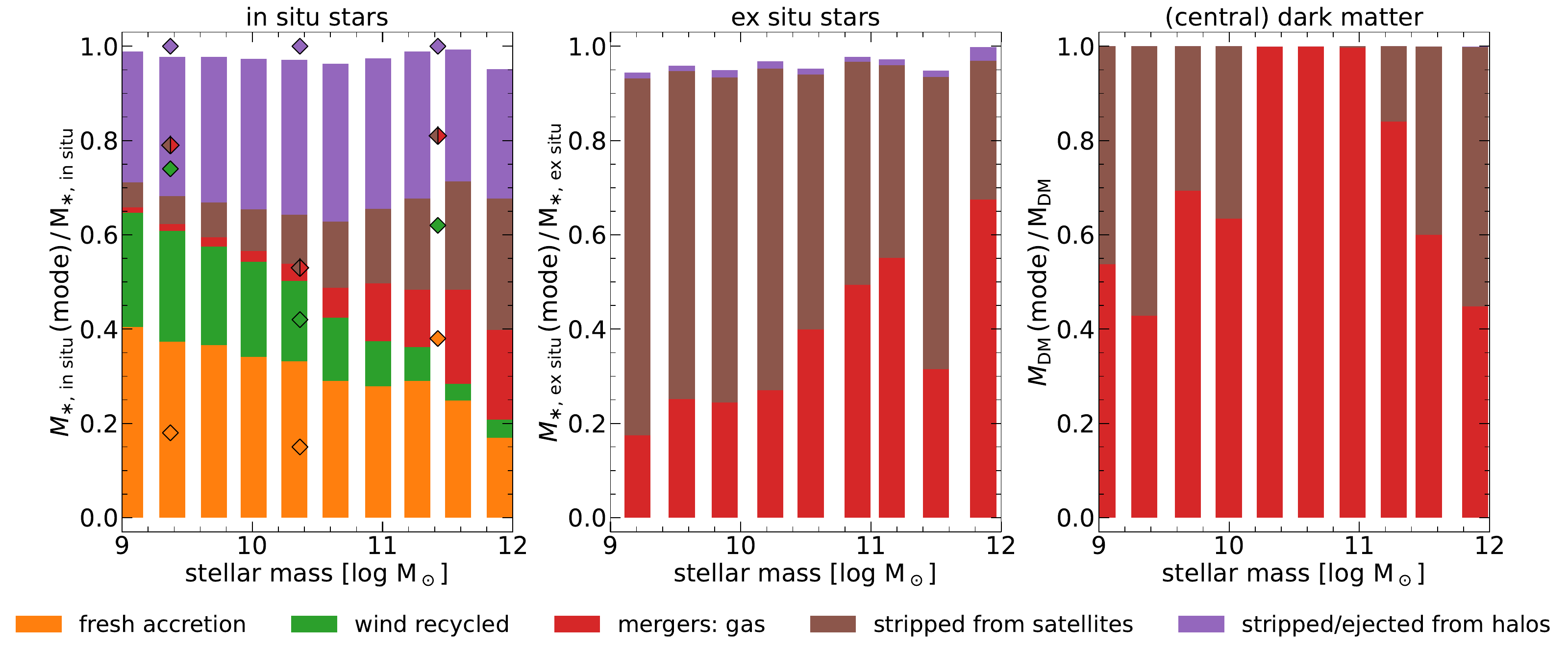}
   \caption{Fraction of the ($z$\,$=$\,$0$) in situ stellar mass of galaxies brought in by distinct accretion channels (left panel). We define fresh accretion as gas acquired from the IGM that has remained in the halo since it first entered (orange). Gas that was recycled between the galaxy and the halo by feedback processes is labeled wind recycled (green). Deposition of gas by satellites directly into the gaseous disks of centrals ($\leq$\,$2\,$\,R$_{\rm SF, 1/2}$; red) is further contrasted against the stripping of gas at farther distances (brown). Lastly, gas that is ejected or stripped from other halos, i.e., not satellites, is shown in purple. Colored diamonds show analogous measurements from the FIRE cosmological simulations \protect{\citep{angalc17}}. Both the bars and the diamonds show stacked values. For comparison, the middle and right plots show the same categories for ex situ stars and dark matter particles, respectively.}
   \label{fig:accretion_channels}
\end{figure*}

The trend of the mean in situ stellar age (red) largely follows that of the in situ formation time computed from all tracers (solid blue) but is vertically offset to lower redshift. That is, although a high fraction of gas is accreted across the virial radius at early epochs, it forms stars following a time delay of several Gyr. We speculate that this is largely due to gas circulation through the halo \citep{angalc17,suresh19}, as motivated in Fig.~\ref{fig:rad_prof_sub}. The in situ (red) and total (orange) stellar age profiles are similar for low-mass halos (M$_{\rm{200c}}$\,$\lesssim$\,$10^{12.5}$\,M$_\odot$), as expected since these are dominated by in situ stars. The mean stellar age of more massive galaxies is higher, implying that the accreted population of ex situ stars is generally older than those produced in situ \citep{lackner12}.

These trends are broadly consistent with \citet{hill17} using observational data, who report mean stellar ages of $z_\star$\,$\sim$\,$[2.08,\,1.49,\,0.82,\,0.37]$ for galaxies of mass M$_\star$\,$\sim$\,$[10^{11.5},\,10^{11.0},\,10^{10.5},\,10^{10.0}]$\,M$_\odot$ at $z$\,$\sim$\,$0.1$. They furthermore find that the difference between $z_\star$ and the stellar assembly time is larger for more massive galaxies, suggesting that the ex situ population in high-mass galaxies is typically older than its in situ counterpart. Although such a relation between $z_\star$ and M$_\star$ is absent in the EAGLE cosmological simulation \citep{hill17}, some SAMs can produce a similar qualitative trend \citep[e.g.,][]{henriques15}.

Recent observations suggest that massive elliptical galaxies form their stellar content at high redshifts, while their low-mass counterparts have late epochs of star formation activity -- galaxy "downsizing" \citep{cowie96}. This agrees with the qualitative stellar age trends in Fig. ~\ref{fig:halo_insitu_formation_times}, which may partly be driven by merger-triggered star formation at lower redshifts decreasing overall in situ stellar ages \citep{thomas10}.

\subsubsection{Accretion channels}\label{subsubsec:accretion_channels}

We proceed to address better the origin vs dynamics of gas accretion and its impact on in situ star formation. Fig.~\ref{fig:accretion_channels} quantifies the mass fraction of gas accreted through five channels, as defined in Section \ref{subsec:gas_accretion_modes}. Fresh accretion is shown in orange and gas recycled in the halo due to winds is displayed in green. We furthermore contrast gas accreted through mergers (red) and gas stripped from infalling satellites (brown). Lastly, we assess gas stripped (or ejected) from other halos in purple. The vertical bars show the results from TNG50,\footnote{As we depict the median value of all galaxies in a stellar mass bin for each category, the sum of all accretion channels does not necessarily add up to one.} while the diamonds contrast against similar measurements from individual galaxies of the FIRE simulations \citep{angalc17}.

Fresh accretion of IGM gas and the transfer of gas in between halos are the two dominant channels for low-mass galaxies (M$_\star$\,$\lesssim$\,$10^{10}$\,M$_\odot$). The total contribution by merging satellites is typically small at these mass ranges, adding up to $f_{\rm bar}$\,$\sim$\,$0.15$, but grows increasingly more important for more massive galaxies, reaching mass fractions of $\sim$\,$0.5$ at M$_\star$\,$\sim$\,$10^{12}$\,M$_\odot$\footnote{We note that this also includes gas stripped during flyby events, in case the accretion onto the central subhalo occurred from one snapshot to the next.}.  Gas stripped or ejected from other halos is equally important at all galaxy stellar masses ($f_{\rm bar}$\,$\sim$\,$0.30$). Moreover, it is the dominant accretion channel for galaxies above $10^{10.5}\,\mathrm{M}_\odot$ in stellar mass. In comparison to FIRE, our TNG50 findings are in reasonable agreement for low-mass galaxies, while values of other channels deviate by $\lesssim$\,$20$\,\%. 

Lastly, we also contrast against the dominant accretion channels for ex situ stars (middle plot) and dark matter (right plot). For the latter, we only consider particles within $2\,$R$_{0.5,\star}$ at $z=0$, that is the DM core. Both these components are primarily stripped from infalling satellites or accrete via merger events, in broad agreement with \citet{madau08} and \citet{angulowhite10}. However, \citealt{genel10} find that mergers contribute only $\lesssim$\,$60\%$ to the total DM mass of M$_{\rm{200c}}$\,$\sim$\,$10^{14}$\,M$_\odot$ halos, while the remaining $\sim$\,$40$\,\% is accreted smoothly. For lower halo masses, they find that the contribution by mergers may approach zero.

As a caveat, all three studies above quantify the build-up of the entire DM halo, while we focus on the core. The high resolution of TNG50 and its ability to resolve smaller substructures may also play a role. In addition, the boundary of $2\,$R$_{\rm SF, 1/2}$ to distinguish merger versus stripped contributions from satellites is somewhat arbitrary and changing this value shifts our results. Finally, we remark that some of these categorizations are sensitive to the snapshot spacing in the simulation. For example, our "stripped or ejected from halos" definition requires that tracers are identified in the IGM at only one (or more) snapshots. 

\begin{figure}
   \centering
   \includegraphics[width=0.48\textwidth]{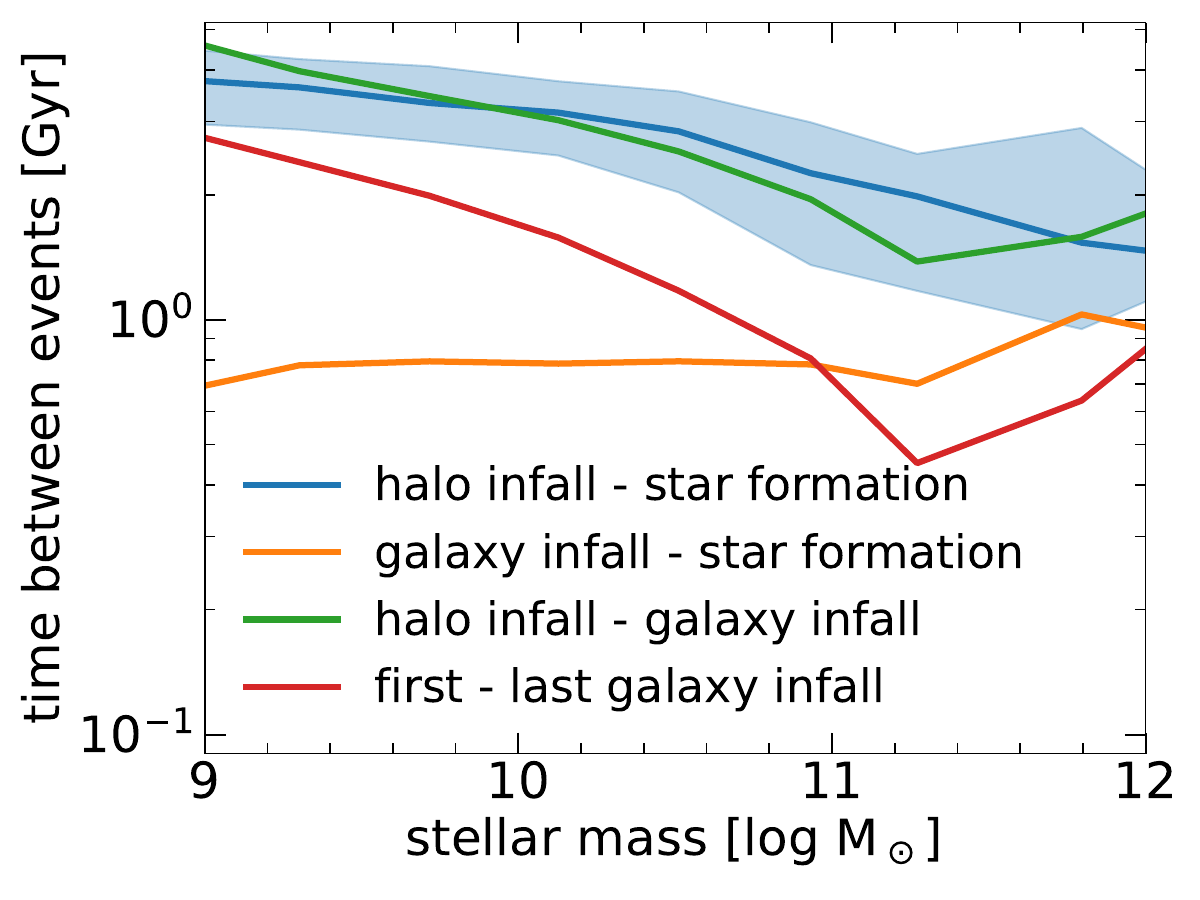}
   \caption{Temporal lag between various spatial crossing and star formation events. In blue (orange), we show the time difference between the moment that a tracer last crossed R$_{\rm{200c}}$ ($2$\,R$_{\rm{SF, 1/2}}$), i.e., halo infall (galaxy infall) and that when its parent gas cell is converted into a star particle. The time gap between the two infall events (green) is contrasted against the time elapsed between the first and the last crossing of $2$\,R$_{\rm{SF, 1/2}}$ (red). All curves show medians across the galaxy population and the value for each galaxy is also the median of all its tracers that actually undergo the event(s), always taking absolute values (see text). Gas accretion onto galaxies proceeds over several $\sim$\,Gyr, while the eventual collapse of ISM gas into stars occurs on shorter time scales ($\lesssim$\,Gyr), as long as it is not recycled back into the halo due to feedback.}
   \label{fig:specific_times}
\end{figure}

\subsubsection{Infall times and star formation}
\label{subsubsec:specific_times}

In Fig.~\ref{fig:specific_times} we quantify the timescales of accreting gas. We measure the median time difference between the moment that a tracer last crossed R$_{\rm{200c}}$ and when it eventually converts into a star particle. This is the halo infall time (blue, shaded area showing galaxy-to-galaxy scatter). Similarly, the time difference between the last crossing of $2$\,R$_{\rm{SF, 1/2}}$ and the moment of star formation is the galaxy infall time (orange). The time gap between the two infall or crossing events is shown in green. The red line depicts the time elapsed between the first and the last crossing of $2$\,R$_{\rm{SF, 1/2}}$. Since stars may form in the halo before they infall into the galaxy (Fig.~\ref{fig:insitu_profile_and_true_false}), that is migrate, and hence have negative galaxy infall-star formation times, we show absolute values on the y-axis. However, the fraction of stars with these properties is low ($\lesssim\,10\%$, see Fig. ~\ref{fig:insitu_profile_and_true_false} right panel) and their inclusion does not qualitatively change our findings. 

As expected, the time gap between (last) galaxy infall and star formation is the shortest. Interestingly, it is largely independent of galaxy stellar mass, varying between $\sim$\,$700$\,$-$\,$900$\,Myr across the mass range shown. On the other hand, the time between (last) halo infall and star formation is significantly higher, ranging from $\sim$\,$5$\,Gyr for M$_\star$\,$\sim$\,$10^{9}$\,M$_\odot$ galaxies to $\sim$\,$2$\,Gyr for their more massive counterparts (M$_\star$\,$\gtrsim$\,$10^{11}$\,M$_\odot$). These numbers are broadly consistent with \citet{birnboim07} and \citet{noguchi18}, who report time scales of $\sim$\,$2$\,$-$\,$3$\,Gyr for M$_{\rm{200c}}$\,$\sim$\,$10^{12}$\,M$_\odot$ halos.

The nominal scatter is several hundred Myr, up to one to two Gyr, across the mass range. Scatter in the galaxy infall - star formation time is less than in the halo - galaxy infall time. This may result from the contribution of stars formed in the halo, increasing the variation for the halo infall-galaxy infall time gap.

This declining trend with mass is largely driven by the time taken by gas to (last) infall into the galaxy following its (last) accretion into the halo (green). This reflects the efficient circulation of gas within low-mass halos by stellar feedback, as shown by the red curve \citep[see also][]{nelson19a,smith19}. Stronger SMBH-driven feedback episodes in higher-mass galaxies typically eject gas out of halos \citep{borrow20,gebhardt24}, resetting the counter for halo infall and leading to smaller differences between subsequent star formation. Baryonic feedback processes clearly play an important role in regulating the efficiency and time delays between cosmic gas accretion and in situ stellar mass buildup.

\begin{figure}
   \centering
   \includegraphics[width=0.48\textwidth]{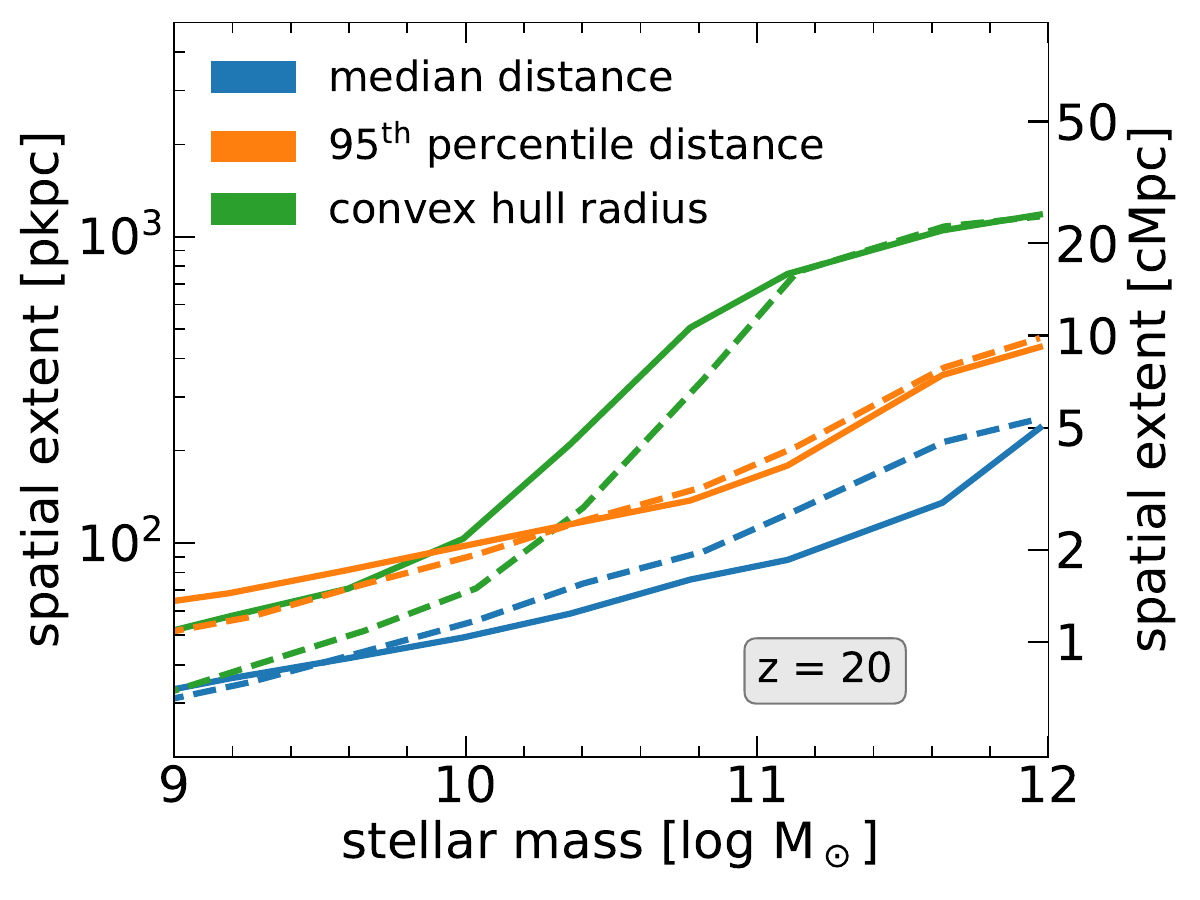}
   \caption{Spatial extent of the initial ($z$\,$\sim$\,$20$) gas distribution that fuels the assembly of in situ (solid curves) and ex situ (dashed) stars by $z$\,$=$\,$0$. We offer three different measurements: the median distance of tracers (blue), the 95$^{\rm{th}}$ percentile distance (orange), and the radius of the sphere whose volume equals that of the convex hull of the initial gas distribution (green). As expected, the collapse of gas into more massive halos takes place from initially larger scales.}
   \label{fig:lagrangian_region}
\end{figure}

\subsubsection{Baryonic Lagrangian regions}\label{subsubsec:bar_lagr_reg}

We have previously seen that the baryonic mass contributing to the ($z$\,$\sim$\,$20$) in situ stellar mass of galaxies comes from large distances and over long timescales. We therefore conclude our analysis in Fig.~\ref{fig:lagrangian_region} by quantifying the spatial extent of the initial ($z$\,$\sim$\,$20$) gas distribution that fuels the assembly of in situ (solid curves) and ex situ (dashed) stars by $z$\,$=$\,$0$. In each case, we make three different measurements: the median distance of tracers (blue), the 95$^{\rm{th}}$ percentile distance (orange), and the radius of the sphere whose volume equals that of the convex hull of the initial gas distribution (green).\footnote{Distances were always computed with respect to the position of the galaxy extrapolated to $z=20$ if it is not present at that time.}

For in situ stars, the median distance increases from $\sim$\,$30$\,pkpc for M$_\star$\,$\sim$\,$10^{9}$\,M$_\odot$ galaxies to $\gtrsim$\,$200$\,pkpc for the most massive objects in TNG50 (M$_\star$\,$\sim$\,$10^{12}$\,M$_\odot$). This corresponds to a large cosmic region of $\sim 5$ comoving Mpc. The 95$^{\rm{th}}$ percentile distance has the same qualitative trend but is offset to larger values by a factor of $\sim$\,$2$ at all stellar masses. The collapse of gas into more massive halos thus takes place from initially larger scales, as expected.

The mass trend of the convex hull-derived Lagrangian region extent is different. These radii are largely similar to the 95$^{\rm{th}}$ percentile distance for M$_\star$\,$\lesssim$\,$10^{10}$\,M$_\odot$ galaxies, but increase sharply by a factor of $\sim$\,$3-5$ toward more massive halos. The morphology of the initial gas distribution of massive galaxies is thus complex and deviates from simple spherical geometries.

The median distance of ex situ stars (dashed) is slightly larger than that of in situ stars (solid) at almost all stellar masses. However, the inverse holds for the radii derived from the convex hulls. This implies that gas that collapses down to in situ stars fills a relatively larger volume at early times, but is more concentrated in the central regions, leading to smaller median values. 

We can compare these baryonic Lagrangian regions to the more commonly studied Lagrangian regions of dark matter halos. In particular, \citet{mowhite96} found Lagrangian regions up to an order of magnitude larger than the values presented here for baryons, albeit at much lower numerical resolution. Focusing on the displacement of baryons from their initial positions, \citet{borrow20} find that nearly half of all baryons move more than $1\,h^{-1}\,\rm{Mpc}$ by $z=0$. This motion is driven by AGN feedback, with some particles traveling distances of up to $12\,h^{-1}\,\rm{Mpc}$. Thus, one would expect baryonic Lagrangian regions to easily span many Mpc as well, potentially with diverse and nonspherical geometries \citep{lee09}.


\section{Summary}\label{sec:summary}

In this work, we use the TNG50 cosmological magnetohydrodynamical simulation \citep{nelson19a,pillepich19} to explore the accretion of the baryons that form the in situ stars of galaxies. In particular, we have used Monte Carlo tracers \citep{genel13} to quantify the cosmological origin of gas that fuels their formation. Our main findings are as follows:

\begin{enumerate}

    \item in situ stars dominate ($\gtrsim$\,$90$\,\%) the stellar population of low-mass galaxies (M$_\star$\,$\lesssim$\,$10^{10}$\,M$_\odot$) while the contribution by ex situ stars grows increasingly important toward the high-mass end. In the most massive galaxies in TNG50, in situ stellar growth ceases by $z$\,$\sim$\,$0.6$, supporting the two-phase stellar mass growth scenario for high-mass halos (Fig.~\ref{fig:ex_in_situ_frac}). At all halo masses, in situ stars are typically centrally concentrated. A high fraction of them form within the central gaseous disk and remain within until $z$\,$=$\,$0$ (Fig.~\ref{fig:insitu_profile_and_true_false}).

    \item Gas forming in situ stars by $z$\,$=$\,$0$ enters high-mass galaxies earlier than their low-mass counterparts. Moreover, the mean infall redshift of gas into the core of galaxies is typically higher than at larger distances (Fig.~\ref{fig:mean_infall_image}).
    
    \item At all halo masses, a dominant ($\gtrsim$\,$90$\,\%) fraction of gas forming in situ stars by $z$\,$=$\,$0$ is present in the IGM at high redshifts ($z$\,$\sim$\,$6$). The remainder is locked in other halos at these early times, ranging from $\sim$\,$5$\,\% for dwarfs to $\sim$\,$10$\,\% for groups (Fig.~\ref{fig:tracer_frac_evo}). Most of the gas from other halos is accreted via mergers and only a low fraction through flybys and/or intergalactic mass transfer (Fig.~\ref{fig:tracer_frac_evo_other_gals})

    \item Groups typically assemble their baryons earlier than their low-mass counterparts. Furthermore, they have smaller Lagrangian half-mass distances (R$_{\rm{L,1/2}}$) at low redshift. The inverse is true at $z$\,$\gtrsim$\,$3$. IGM gas collapses into halos from larger distances than gas contributed by mergers, resulting in larger IGM-R$_{\rm{L,1/2}}$ values (Figs.~\ref{fig:radial_profiles} and \ref{fig:hmr_vs_halo_mass}).

    \item The dark matter halo and its central core assemble earlier for low-mass halos. In contrast, gas that eventually fuels in situ star formation is accreted into high-mass halos at earlier cosmic epochs than their low-mass counterparts. Smooth accretion of IGM gas occurs earlier than clumpy accretion via satellites for high-mass halos (M$_{\rm{200c}}$\,$\gtrsim$\,$10^{11.5}$\,M$_\odot$) while the opposite holds for lower-mass halos (Fig.~\ref{fig:halo_insitu_formation_times}). The contribution of gas from mergers is higher for more massive halos while wind recycling and smooth accretion are lower (Fig.~\ref{fig:accretion_channels}).

    \item Following accretion into the halo, gas is typically recycled between the galaxy and the circumgalactic medium prior to its collapse into (in situ) stars by $z$\,$=$\,$0$. This gives rise to a roughly $\sim 1-2$\,Gyr time offset between halo infall and eventual star formation (Fig.~\ref{fig:specific_times}).

    \item The collapse of gas into more massive halos takes place from initially larger scales. Moreover, the morphology of the initial Lagrangian gas distribution is less spherical for massive halos, reaching $\sim$\,10s of comoving Mpc (Fig.~\ref{fig:lagrangian_region}).
\end{enumerate}

This work presents an initial exploration into the cosmological origin of gas that fuels in situ star formation. Clear avenues for future explorations exist. For instance, the current analysis is limited at the high-mass end to objects of M$_{\rm{200c}}$\,$\sim$\,$10^{14}$\,M$_\odot$. Utilizing cosmological simulations run with larger boxes would make it possible to understand and quantify the in situ stellar build-up of more massive halos, that is clusters, as well as other galactic components of interest, such as bars, bulge, and central supermassive black holes.

\section*{Data availability}

The IllustrisTNG simulations including TNG50 are publicly available and accessible at \url{www.tng-project.org/data} \citep{nelson19b}. Data from this publication is available on request from the corresponding authors. This work has benefited from the \texttt{scida} analysis library \citep{byrohl24}.

\begin{acknowledgements}
RR and DN acknowledge funding from the Deutsche Forschungsgemeinschaft (DFG) through an Emmy Noether Research Group (grant number NE 2441/1-1). RR is a Fellow of the International Max Planck Research School for Astronomy and Cosmic Physics at the University of Heidelberg (IMPRS-HD).  
This work has made use of the VERA supercomputer of the Max Planck Institute for Astronomy (MPIA).
\end{acknowledgements}

\bibliographystyle{aa}
\bibliography{refs}

\clearpage

\vspace{8cm}

\begin{appendix}

\section{Numerical resolution convergence} \label{sec_convergence}

\begin{figure}[h!]
   \centering
   \includegraphics[width=0.48\textwidth]{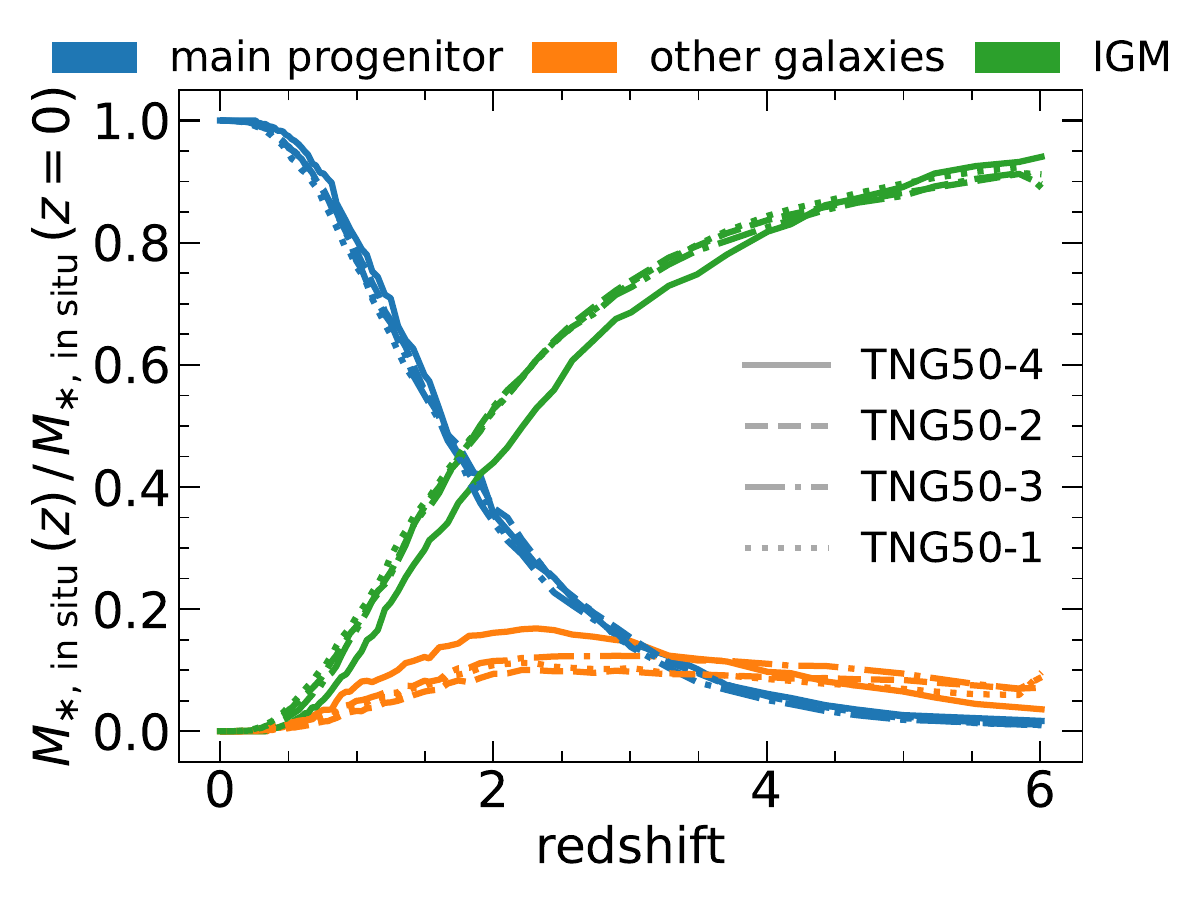}
   \caption{Comparison of Fig. \ref{fig:tracer_frac_evo} for different TNG50 resolution runs. The solid line indicates TNG50-4, the dash-dotted line TNG50-3, the dashed line TNG50-2, and the dotted line TNG50-1. The blue color represents the in situ stellar mass fraction of Milky Way-like central galaxies for tracers in the main progenitor, orange is the mass fraction for tracers in other galaxies, and green is the mass fraction for tracers in the IGM.}
   \label{fig:tracer_frac_res_comp}
\end{figure}

\begin{figure}[h!]
   \centering
   \includegraphics[width=0.48\textwidth]{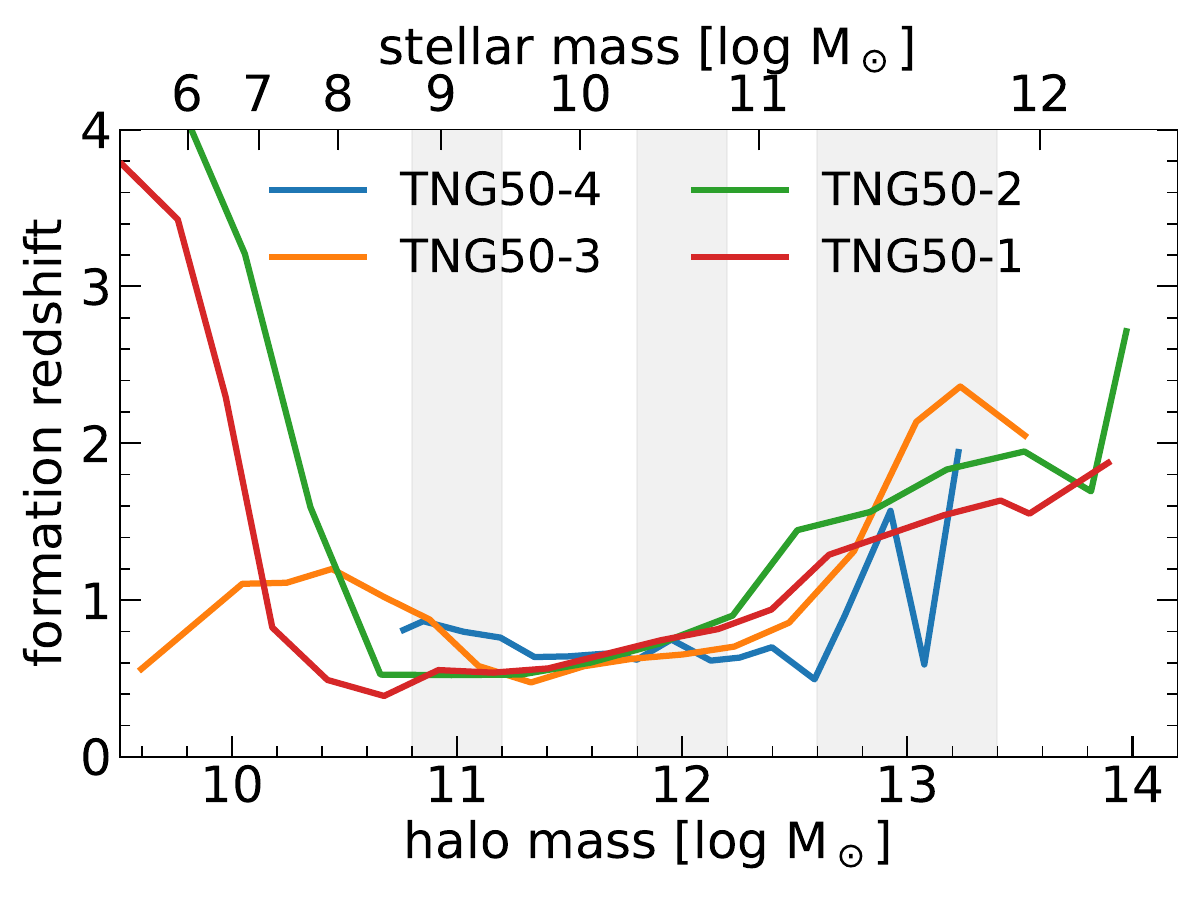}
   \caption{Comparison of galaxy age, i.e., the median age of the stellar content of a galaxy, (compare Fig. \ref{fig:halo_insitu_formation_times}) for different TNG50 resolution runs. The red line indicates TNG50-1 with the highest resolution, green stands for TNG50-2, orange is for TNG50-3, and the blue line represents TNG50-4 with the lowest resolution. The figure spans to lower masses to better visualize resolution effects.}
   \label{fig:gal_age_res_comp}
\end{figure}

\begin{figure}[h!]
   \centering
   \includegraphics[width=0.48\textwidth]{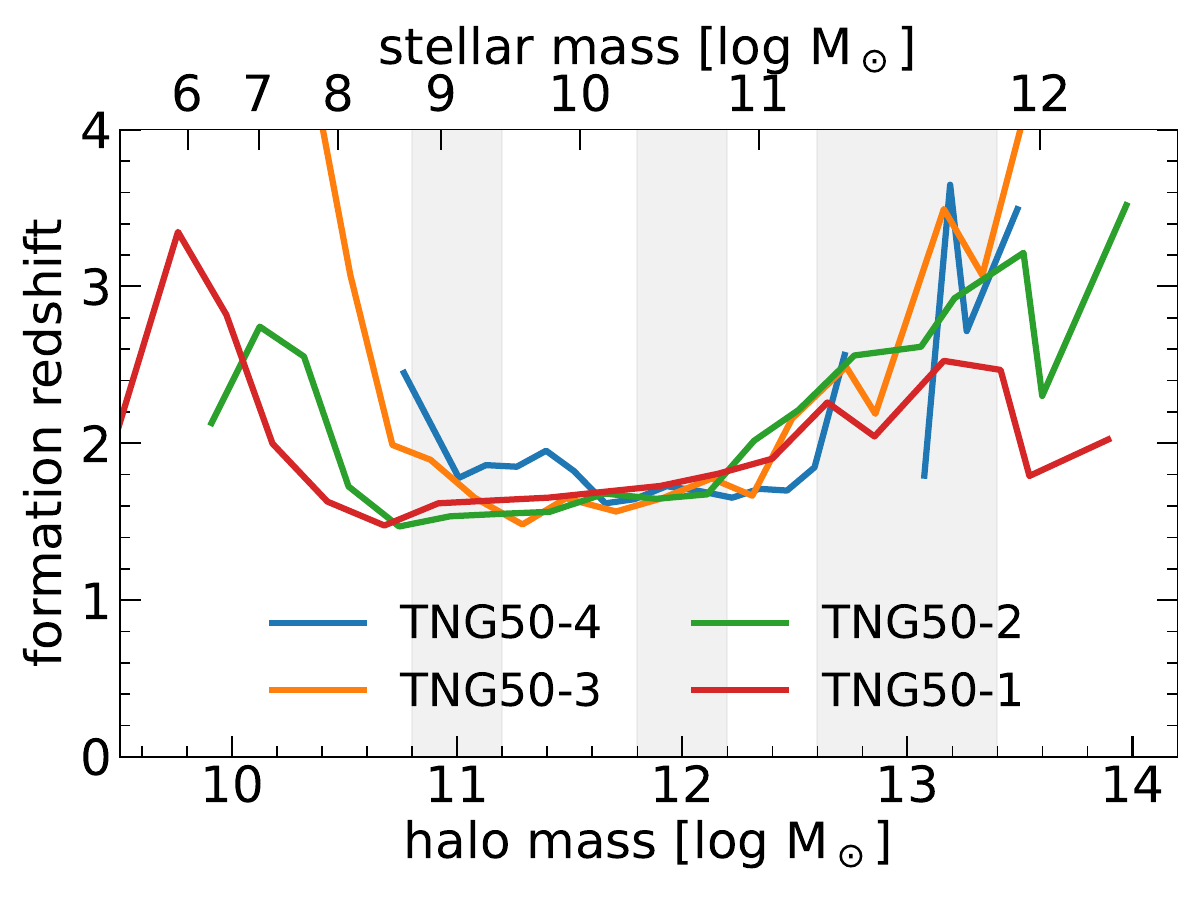}
   \caption{Comparison of the in situ stellar content (compare Fig. \ref{fig:halo_insitu_formation_times}) for different TNG50 resolution runs. The red line indicates TNG50-1 with the highest resolution, green stands for TNG50-2, orange is for TNG50-3, and the blue line represents TNG50-4 with the lowest resolution. The figure spans to lower masses to better visualize resolution effects.}
   \label{fig:insitu_z_res_comp}
\end{figure}

In Fig. \ref{fig:tracer_frac_res_comp} we show the in situ stellar mass fractions in the main progenitor (blue), IGM (green), and other galaxies (orange). In contrast to Fig. \ref{fig:tracer_frac_evo}, the different line styles indicate the resolution levels of TNG50, with solid being the lowest resolution (TNG50-4) and dotted lines the highest (TNG50-1). Here we show the Milky Way mass galaxy subsample.

Apart from the significant change when increasing the resolution from TNG50-4 to TNG50-3, the results become invariant to resolution beyond this point. The difference at low resolutions shifts some tracers (up to 10\% of the total number) from the IGM to other galaxies, possibly due to gas at the border between halos and the IGM. The main progenitor category is mostly unchanged. Overall, our results are impressively well converged across all three resolutions of the IllustrisTNG boxes, especially from TNG50-2 to TNG50-1.

We note that there is significantly more scatter for group-sized centrals, due to their low number statistics in TNG50. As a result, resolution convergence is less clear. Considering dwarfs, there is a stronger resolution trend. Here, lower resolutions may result in only a handful or even one single in situ star tracer. That is, dwarfs below a given mass are simply not resolved in the lower resolution boxes.

Figure \ref{fig:gal_age_res_comp} compares the galaxy stellar ages and Figure \ref{fig:insitu_z_res_comp} the in situ stellar content formation times introduced in Fig. \ref{fig:halo_insitu_formation_times} at different resolution levels. For both quantities, the resolution does not change the result from $\sim 10^{10}\,\rm M_\odot$ until $\sim 10^{11}\,\rm M_\odot$. Above this mass range, the trends are noisy, but higher resolution tends to yield lower values. At the lower-mass end, we see an expected shift toward lower halo mass for higher resolution. The steep feature at low-mass marks galaxies with only a few hundred star particles, that is it demarcates the resolution limit of the runs. In our masses ranges of interest, resolution convergence for these formation time metrics is not a concern.

\section{Halo-to-halo variation}\label{sec:halo_var}

\begin{figure*}
   \centering
   \includegraphics[width=0.68\textwidth]{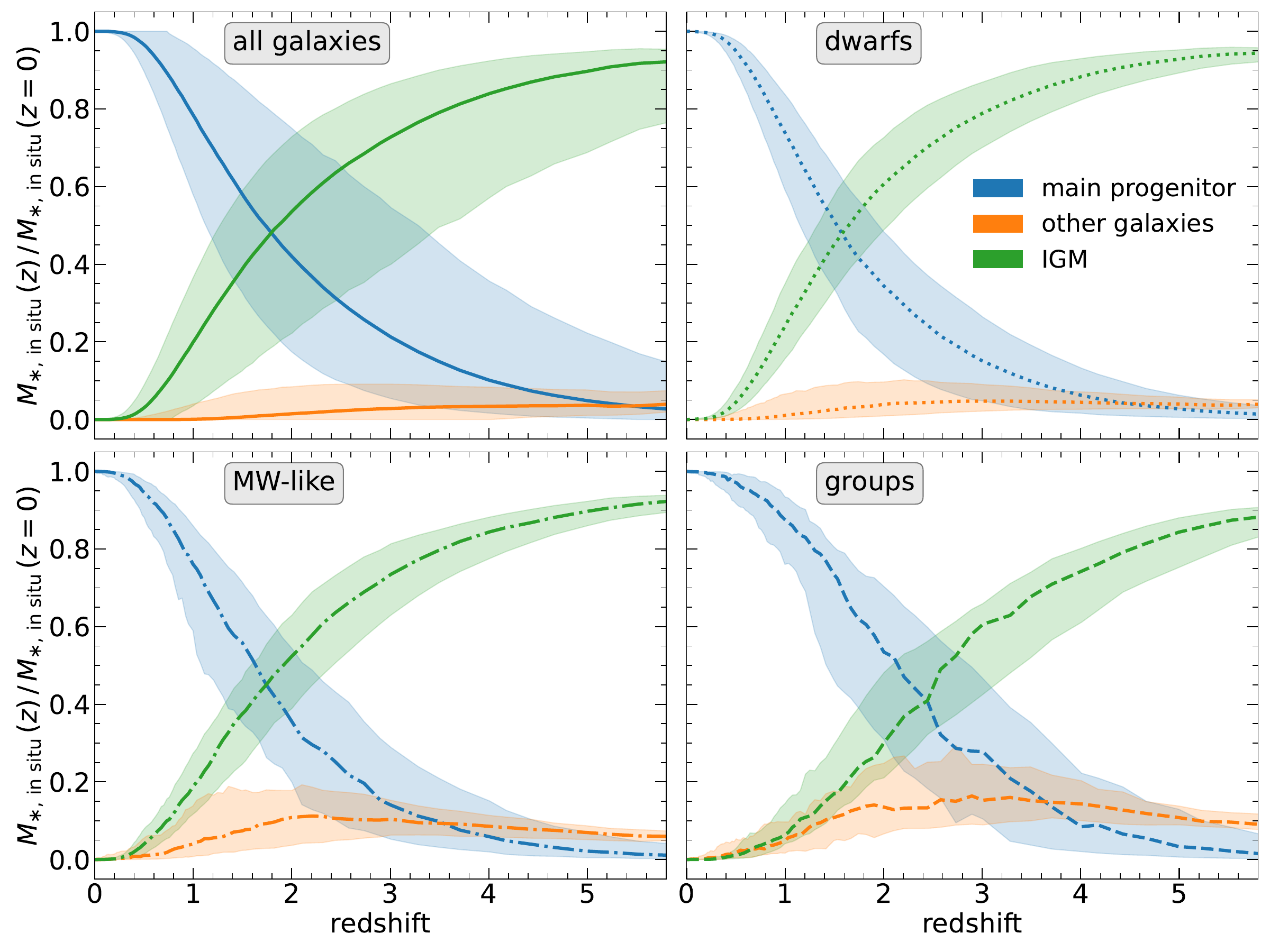}
   \caption{Comparison of Fig. \ref{fig:tracer_frac_evo} for the four different halo mass bins from Table \ref{table:halo_bins} showcasing the halo-to-halo variations discussed in Section \ref{subsubsec:tracer_fracs}. The blue color represents the in situ stellar mass fraction of central galaxies for tracers in the main progenitor, orange is the mass fraction for tracers in other galaxies, and green is the mass fraction for tracers in the IGM.}
   \label{fig:tracer_frac_var_comp}
\end{figure*}

Figure \ref{fig:tracer_frac_var_comp} compares the halo-to-halo variations of the different quantities shown in Fig. \ref{fig:tracer_frac_evo}. The largest scatter is visible for the upper left panel displaying the median lines of all galaxies of the simulation volume, especially for the IGM ($\sim$\,$20\%$) and main progenitor ($\sim$\,$20\%$) tracer fractions. In contrast, the individual halo mass bins show fewer variations in these categories. Generally, the fraction of tracers in the main progenitor varies the most ($\sim$\,$15\%$) and the fraction in other galaxies the least ($\lesssim$\,$10\%$). Considering all galaxies of the simulation volume, the large scatter likely originates from the large number of low-mass objects with little tracers.

\end{appendix}

\end{document}